\documentclass[journal=jacsat,manuscript=article]{achemso}

\usepackage[version=3]{mhchem} % Formula subscripts using \ce{}
\usepackage{multirow}
\usepackage{graphicx}
\usepackage{subcaption}
\usepackage{diagbox}
\usepackage{makecell}
\usepackage{longtable}
\usepackage{caption}
\usepackage{pdfpages}
\author{Yasin Ameslon}\affiliation[HIERN]
{Helmholtz Institute Erlangen-Nürnberg for Renewable Energy, Forschungszentrum Jülich,  Nürnberg, Germany}
\author{Hao Liu}\affiliation[University at Buffalo]
{Department of Materials Design and Innovation, University at Buffalo, NY, USA}\author{Jens Harting}
\affiliation[HIERN]
{Helmholtz Institute Erlangen-Nürnberg for Renewable Energy, Forschungszentrum Jülich,  Nürnberg, Germany}
\alsoaffiliation[FAU]
{Department of Chemical and Biological Engineering and Department of Physics, Friedrich-Alexander-Universität Erlangen-Nürnberg, Erlangen, Germany}
\author{Olivier J.J. Ronsin}\affiliation[HIERN]
{Helmholtz Institute Erlangen-Nürnberg for Renewable Energy, Forschungszentrum Jülich,  Nürnberg, Germany} 
\author{Olga Wodo}\affiliation[University at Buffalo]
{Department of Materials Design and Innovation, University at Buffalo, NY, USA}
\email{olgawodo@buffalo.edu}

\title{Taxonomy of amorphous ternary phase diagrams: the importance of  interaction parameters}


\begin{document}

\begin{abstract}
Understanding phase diagrams is essential for material selection and design, as they provide a comprehensive representation of the thermodynamics of mixtures. This work delivers a broad and systematic overview of possible ternary phase diagrams for amorphous systems representative of polymers, small organic molecules, and solvents. Thanks to computationally efficient methods, an unprecedented library of $>$80,000 ternary phase diagrams is generated based on a systematic screening of interaction parameters. Twenty-one phase diagram types, including unreported ones, are identified. They are classified according to simple rules related to the number of immiscible material pairs, of miscibility gaps, and of three-phase regions. They are mapped onto the three-dimensional interaction parameters space, providing a clear picture of their likelihood and existence conditions. Four well-known phase-diagram types with 0, 1, 2, or 3 immiscible pairs are found to be the most likely. The numerous uncommon phase diagrams are mostly observed within a small parameter window around the critical interaction parameter values. For the most common phase diagram types, we show that the size of the processability window becomes sensitive to interaction parameter variations close to critical values. The sensitivity decreases for materials with increasing molar size. Finally, successful comparisons of simulated and experimental phase diagrams showcase the real-world relevance of this theoretical analysis. The presented results lay a robust foundation for rational design of solution processing conditions and for blend morphology control. Immediate applications include organic thin films and the identification of green solvents for sustainable processing.
\end{abstract}

%%%%%%%%%%%%%%%%%%%%%%%%%%%%%%%%%%%%%%%%%%%%%%%%%%%%%%%%%%%%%%%%%%%%%
%% Start the main part of the manuscript here.
%%%%%%%%%%%%%%%%%%%%%%%%%%%%%%%%%%%%%%%%%%%%%%%%%%%%%%%%%%%%%%%%%%%%%
\section{Introduction}
Understanding the phase behavior of polymer blends and solutions is crucial for designing processing conditions for devices with tailored properties~\cite{paul2012polymer}.
The change of solvent and the inclusion of additives or solvents can regulate the phase behavior by triggering earlier phase separation and crystallization or by affecting the number of phases~\cite{mcdowell2018solvent}.
This opens the opportunity for the rational design of phase behavior through material selection. 
However, with hundreds of possible solvents, choosing solvent blends, additives, and similar strategies represents a combinatorically large set of possibilities. 
Consequently, it has remained challenging to establish reliable design rules for solvent selection in multi-component systems and then further tune the processing conditions based on the choices made~\cite{ethier2024predicting}. 
This paper uses phase diagrams to represent the phase behavior in the material system.
It deals with ternary amorphous systems, for which the phase behavior has already been extensively studied but often with a narrow scope, focusing on specific types of phase diagrams.
Former studies mostly explored the spinodal curves and narrow behavior of the binodal and critical points.
As a typical example, a very recent work~\cite{zhang2024phase} reiterated the need to systematically study phase behavior in a mixture of two solvents with a polymer. 
Zhang~\cite{zhang2024phase} studies the co-nonsolvency effect when a homopolymer chain collapses in a mixture of two good solvents. 
He derived the existence conditions for ternary phase diagrams with closed-loop isolated miscibility gap with two critical points.
The existence conditions have been linked to two mechanisms of phase separation: preferential interaction-driven and solvent-cosolvent attraction-driven mechanisms. 
However, that work focused on one specific type of phase diagram. 

The present contribution takes a different approach to deriving the existence rules of the phase diagram.
Using high-throughput exploration, we first screen the range of interaction parameters, classify the phase diagram to form the taxonomy, and then learn rules for the existence of all the observed phase diagram types and for the sensitivity of the phase diagram type to changes in the interaction parameters. 
The established existence rules are based on the number of immiscible pairs of components. 
Although most of the existence rules are intuitive, the detailed mapping between interaction parameter ranges and the type of phase diagram is less obvious. 
We also report the sensitivity of the miscibility depth of the phase diagrams for three types of material systems (P-SM-S, SM-SM-S, and SM-S-S, whereby P stands for 'polymer' and SM for a 'small molecule' material whose molar volume is, however, significantly larger than the 'solvent' S). 
For some material systems, the depth is relatively insensitive to the interaction parameters, while for other types, the depth becomes very sensitive to the value of the interaction parameter combinations.
Moreover, for material systems located very close to the planes of critical interaction parameters in the 3D-parameter space, the type of phase diagram becomes very sensitive to small changes in interaction parameters. 
In this zone of the design space, additional yet infrequent types of phase diagrams can be found.
Finally, this paper reports the experimental validation for the most typical phase diagrams.
Overall, this work has practical implications for quick solvent selection and co-solvent design in organic thin film fabrication. It provides broad general guidelines for the thermodynamic behavior of ternary amorphous blends. 
Additionally, optimization of solution processing often requires deep understanding and very involved (experimental or theoretical) studies on the kinetic of morphology evolution upon drying. In this context, our findings can also be used to select the most appropriate systems or class of systems based on the class of phase diagram for the subsequent extensive studies. 

%%%%%%%%%%%%%%%%%%%%%%%%%
\section{Method}
First, a large library of ternary phase diagrams is generated. 
The phase diagrams are generated for three materials systems with different combinations of molar sizes and sampled from the ternary space of interaction parameters. 
Our library is then subjected to a comprehensive analysis, employing a combination of techniques to uncover a small number of groups of phase diagrams that share striking similarities regarding the number and distribution of phases within the phase diagrams.
The groups are then mapped back to the input space of interaction parameters to infer existence rules.
This section provides details of the phase diagram construction, high-throughput data generation, and analysis. 

\subsection{Phase diagram construction}
In this study, phase diagrams are generated for ternary blends ($n=3$) described by the Flory Huggins equation for the free energy of mixing~
\cite{ronsin_formation_2022, ronsin_phase-field_2020, ronsin_phase-field_2021, ronsin_phase-field_2022, ronsin_role_2020}:
\begin{equation}
    \label{eq:FloryHuggins}
    \Delta G^V = \frac{RT}{v_0} \left( \sum^n_{i=1}{\frac{\varphi_i ln \varphi_i}{N_i}}+\sum^n_{i=1}{\sum^n_{j>i}{\varphi_i \varphi_j \chi_{ij}}} \right)
\end{equation}
whereby ~$\varphi_{i}$ are the volume fractions of the materials in the mixture. 
The first term in Eq.~\ref{eq:FloryHuggins} has an entropic origin promoting the mixing of the components; the entropy of mixing is positively correlated with the decrease of the molar size $N_{i}$ of a component $i$. The molar size corresponds to the ratio $N_{i}=\frac{v_{i}}{v_{0}}$ between the molar volume $v_{i}$ of a component $i$, and the molar volume $v_{0}$ of the smallest component.

The second term in Eq.~\ref{eq:FloryHuggins} has an enthalpic origin and promotes mixing or demixing, depending on the interaction energy between components $i$ and $j$, the interaction strength being defined through the interaction parameter $\chi_{ij}$. $T$ and $R$ are the temperature and the gas constant, respectively.

Using the equation above, the critical interaction parameter $\chi^c_{ij}$ for a pair of components with molar sizes $N_i$ and $N_j$ is found to be
\begin{equation}
    \label{eq:CriticalInteractionParameter}
    \chi_{ij}^{c} = \frac{1}{2}\left( \frac{1}{\sqrt{N_{i}}}+\frac{1}{\sqrt{N_{j}}}\right)^{2}
\end{equation}
A miscibility gap is expected if the interaction parameter is above the critical value for a given pair of components. 
The critical interaction parameter ranges from nearly zero (for long polymers $N_1$ and $N_2>1000$) to two (for two solvents with $N_1=N_2=1$).
This quantity is defined for a pair of components, and in this paper, it is used to derive the existence rules for the ternary material systems.

\subsubsection{Convex hull approach for the determination of the number of phases}
\begin{figure}[H]
        \includegraphics[width = 0.25\textwidth,trim={9cm 0 0 7cm},clip]{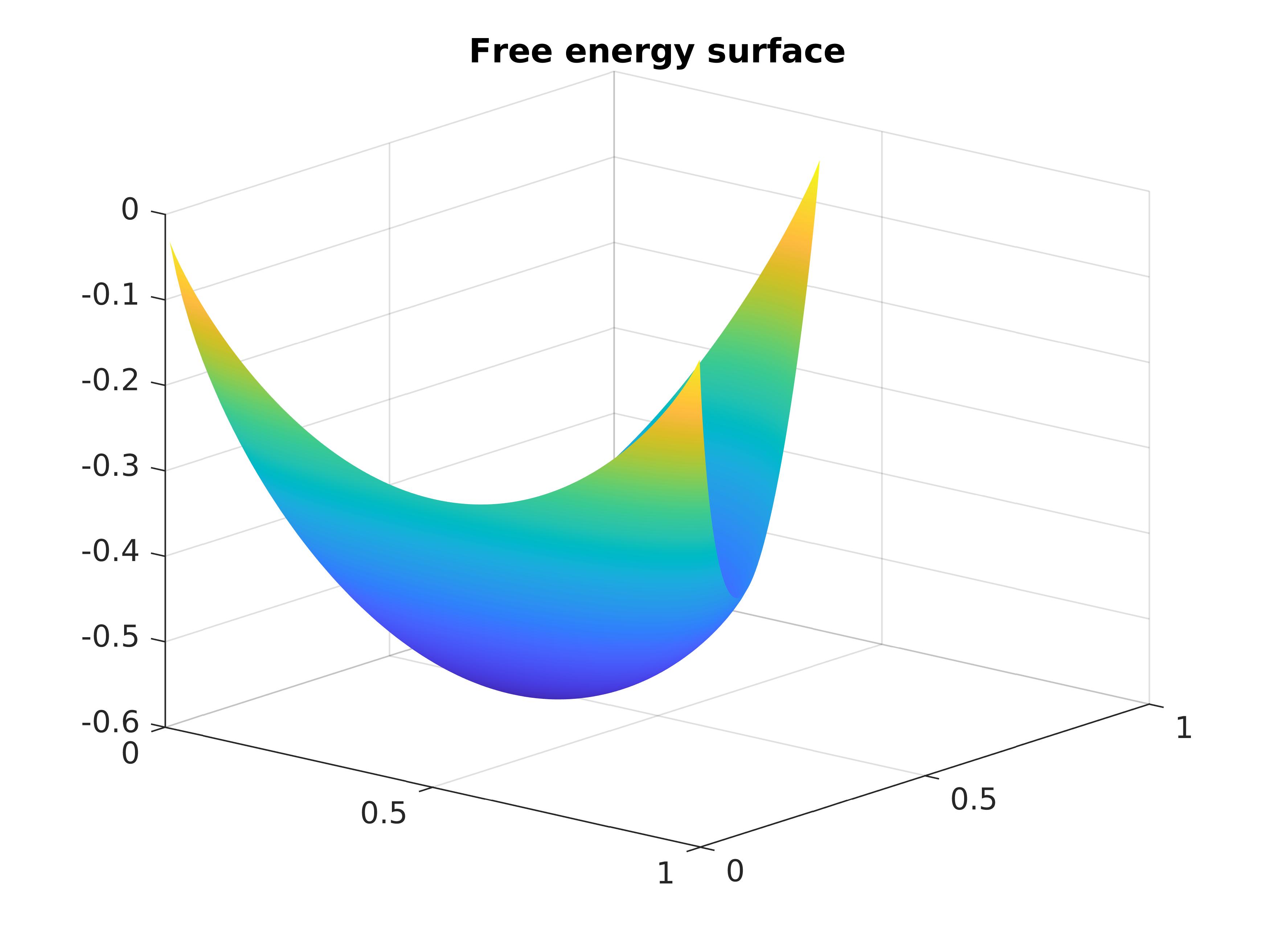}
        \includegraphics[width = 0.25\textwidth,trim={9cm 0 0 7cm},clip]{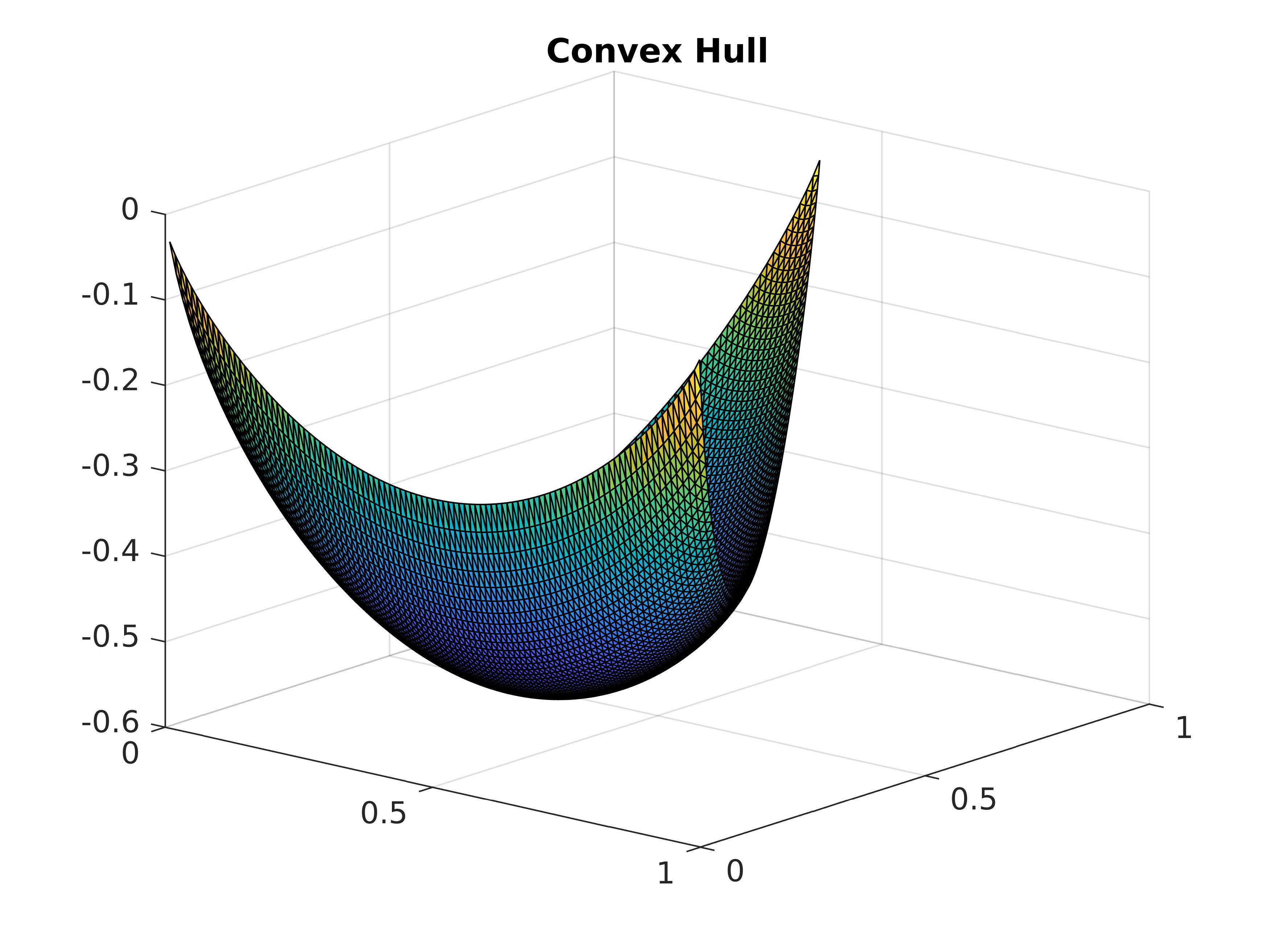}
        \includegraphics[width = 0.2\textwidth,trim={0cm 0 0 6.6cm},clip]{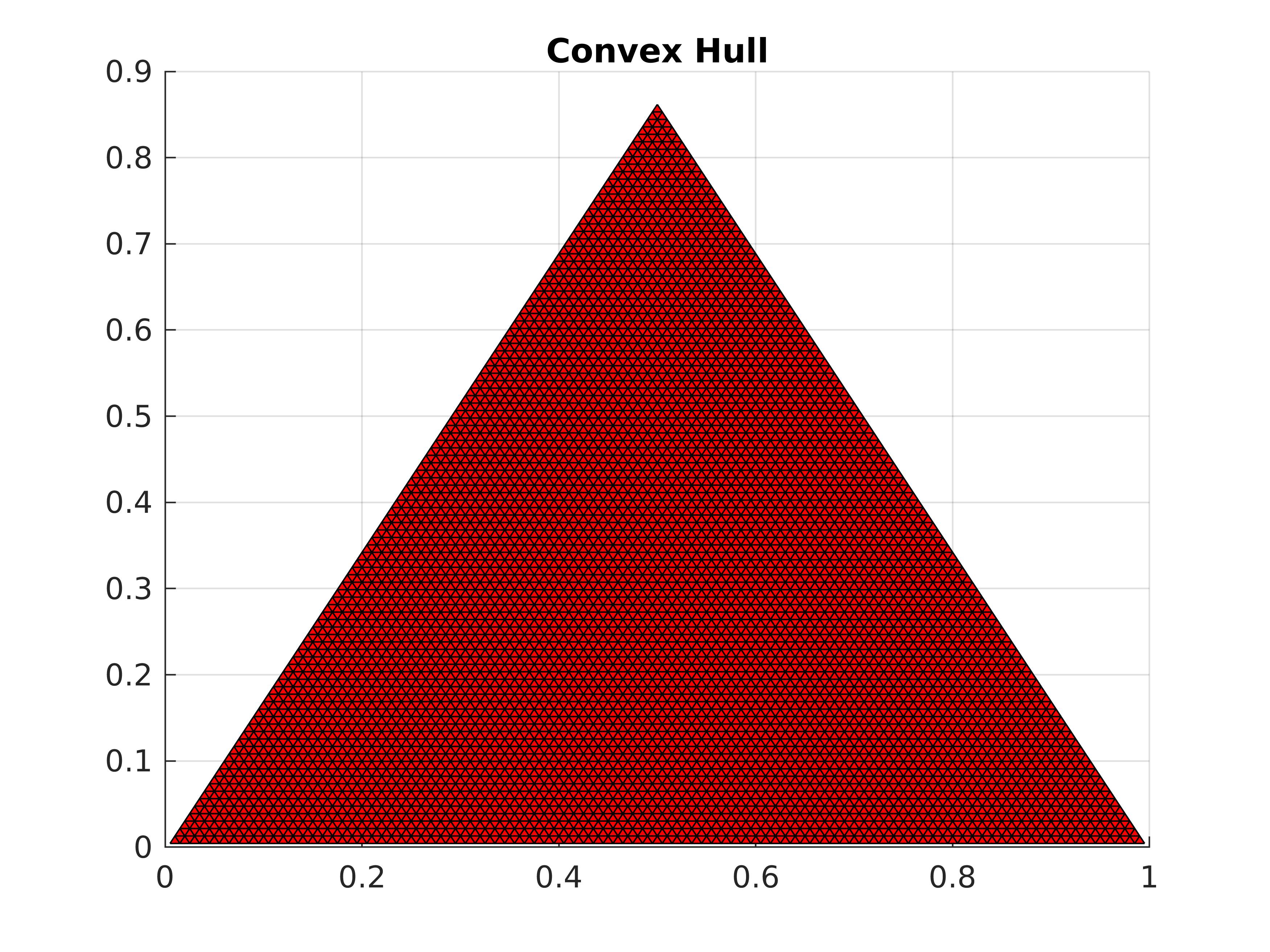}
        \includegraphics[width = 0.22\textwidth]{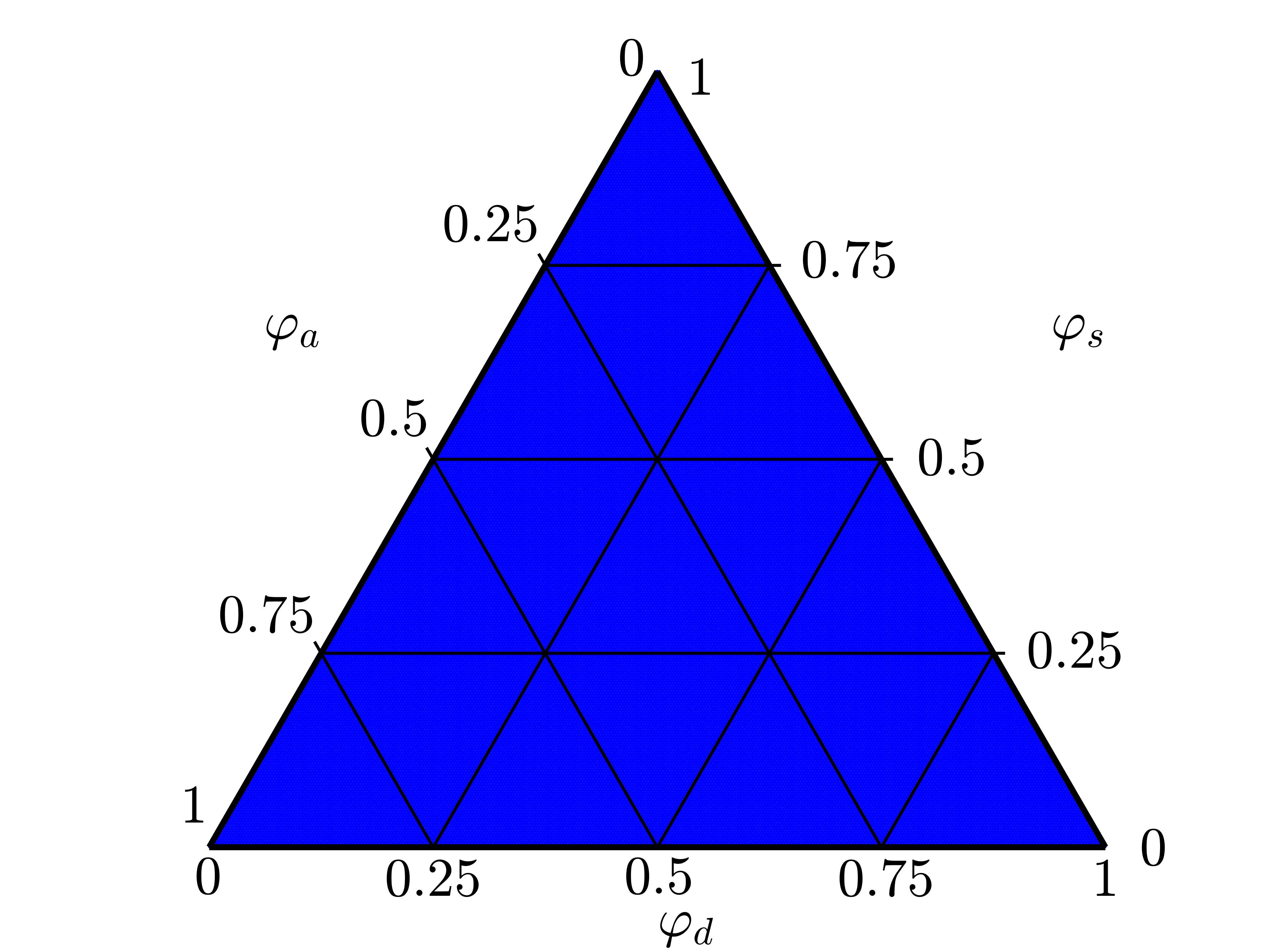} \\
        \includegraphics[width = 0.25\textwidth,trim={9cm 0 0 7cm},clip]{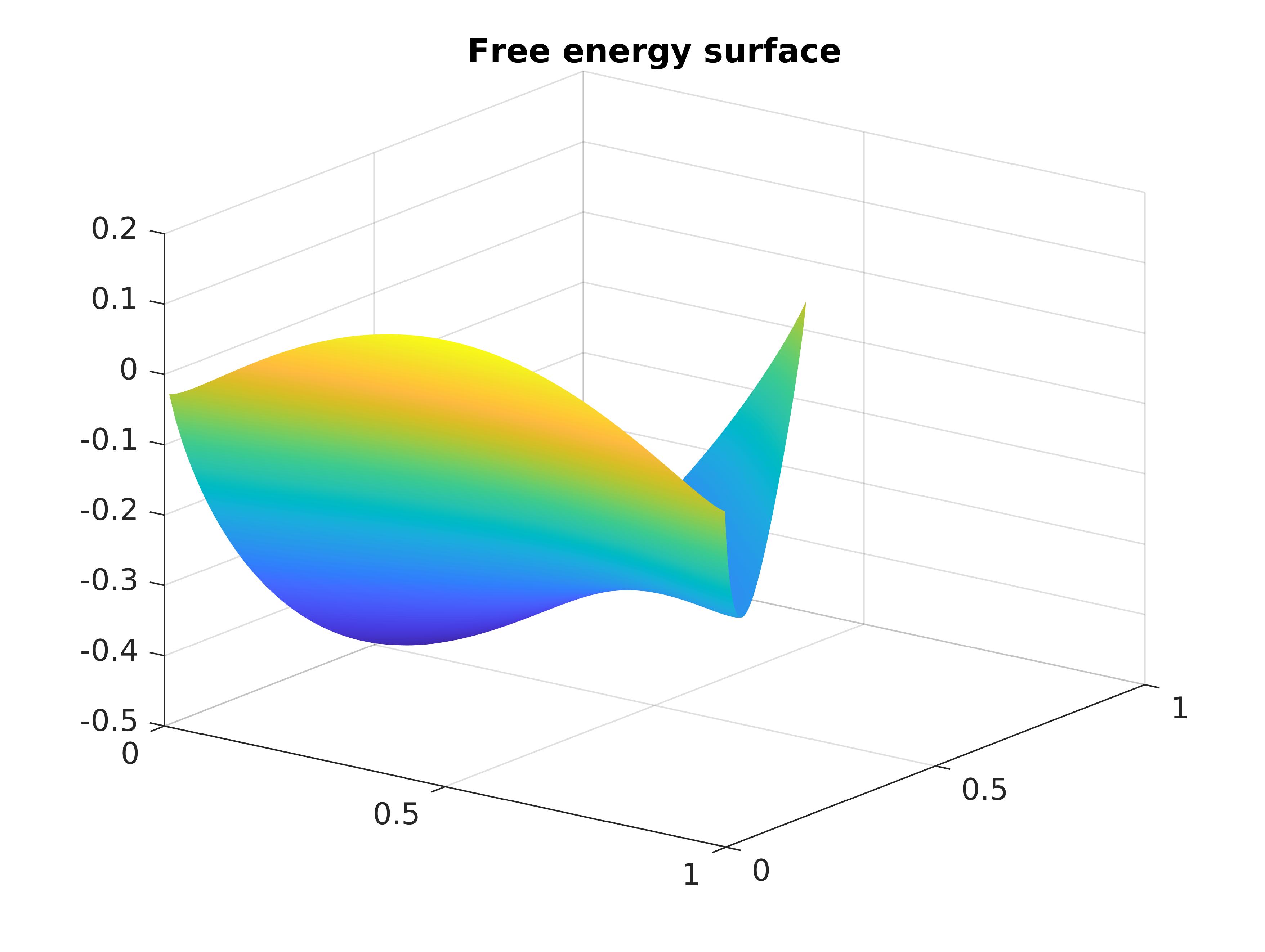}
        \includegraphics[width = 0.25\textwidth,trim={9cm 0 0 7cm},clip]{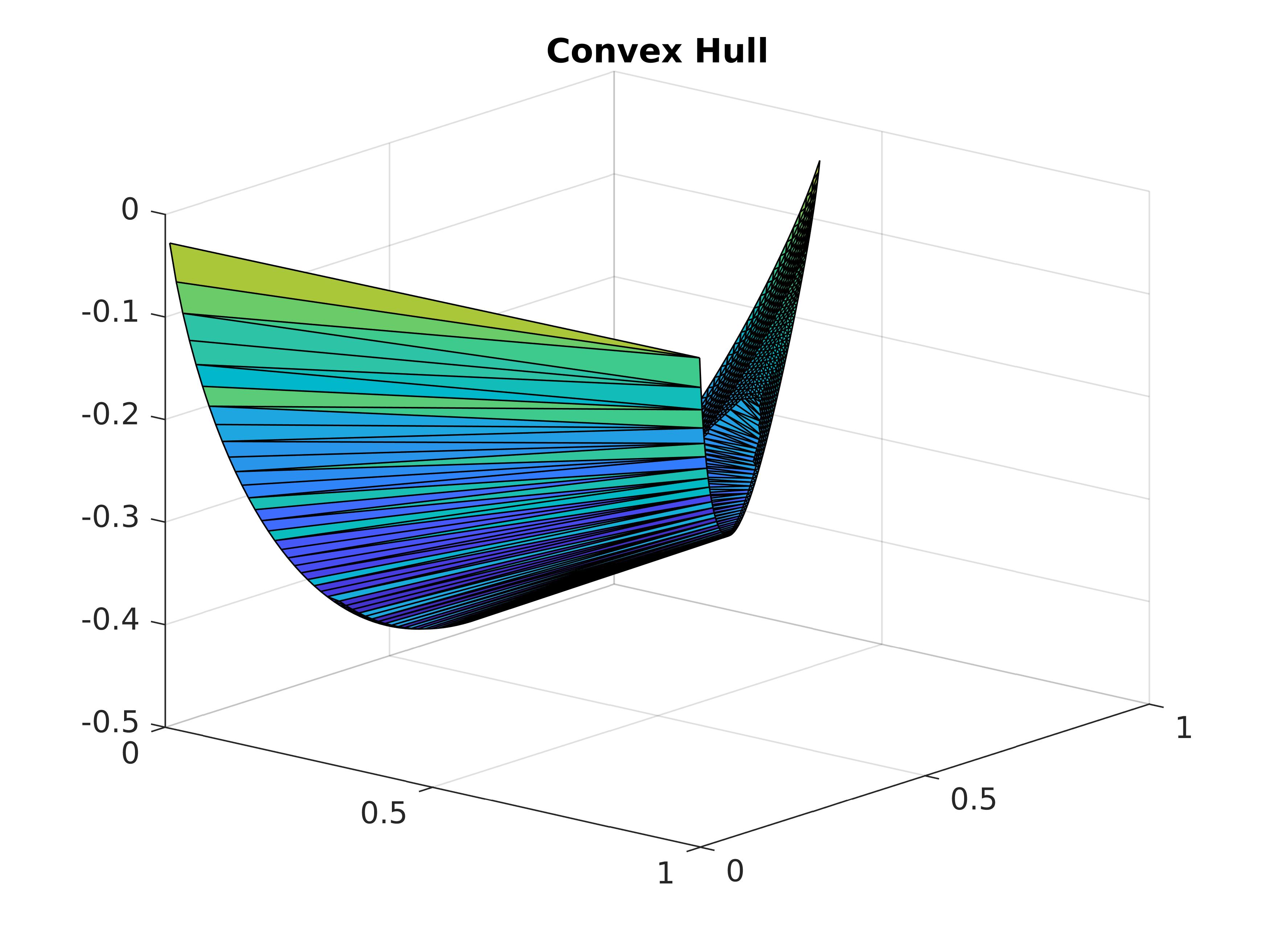}
        \includegraphics[width = 0.2\textwidth,trim={0cm 0 0 6.6cm},clip]{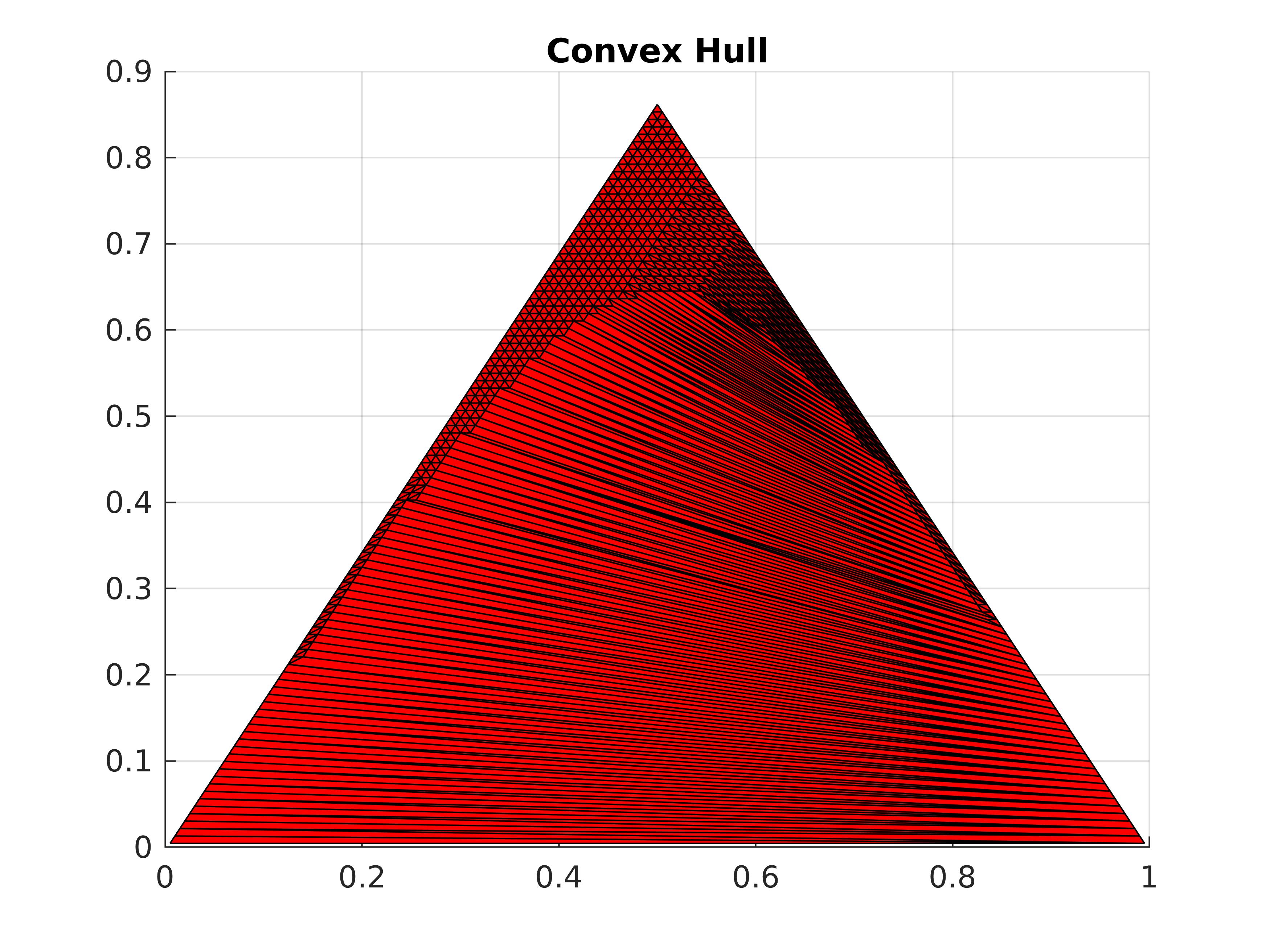}
        \includegraphics[width = 0.22\textwidth]{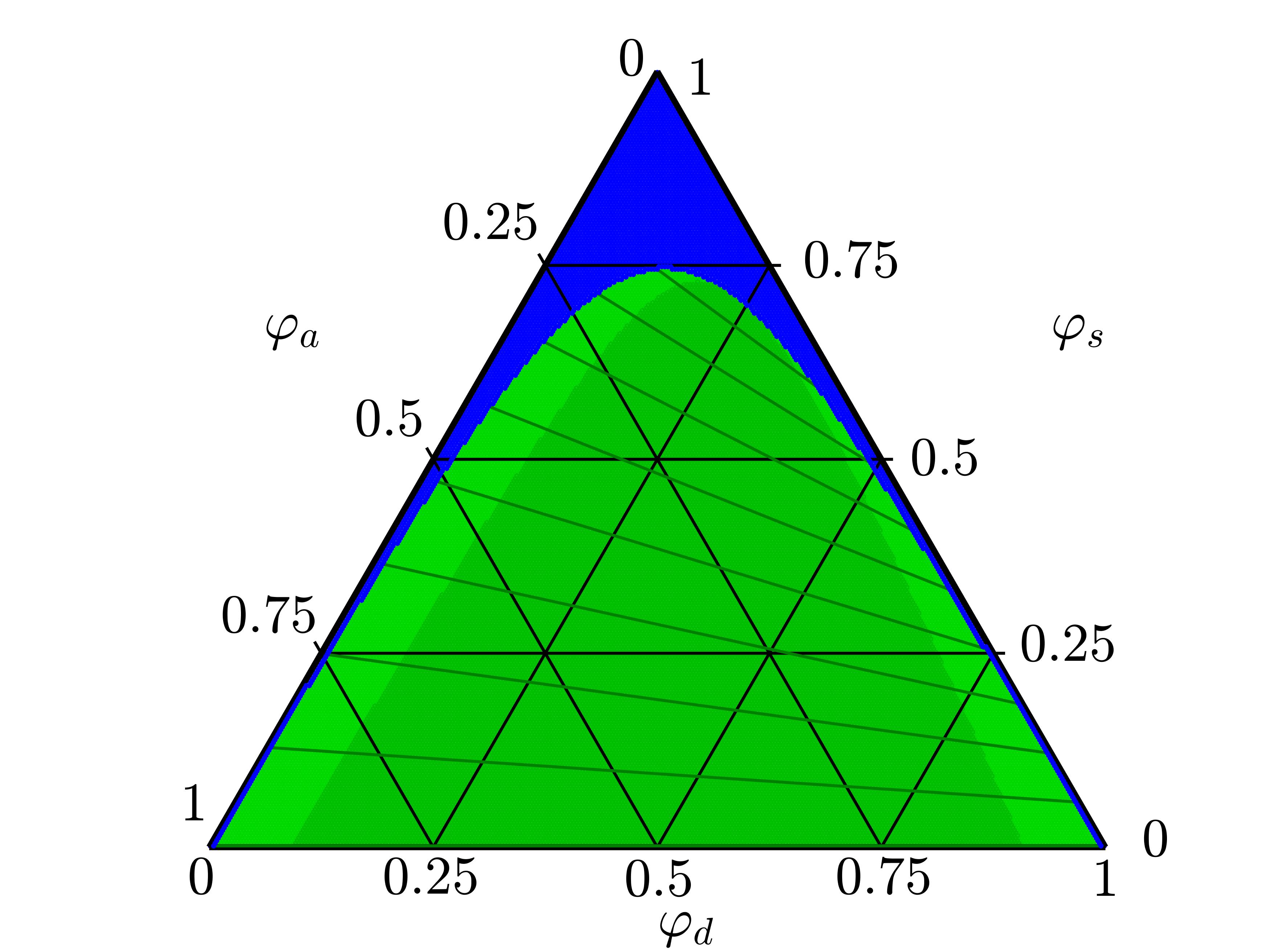}  \\
        \includegraphics[width = 0.25\textwidth,trim={9cm 0 0 7cm},clip]{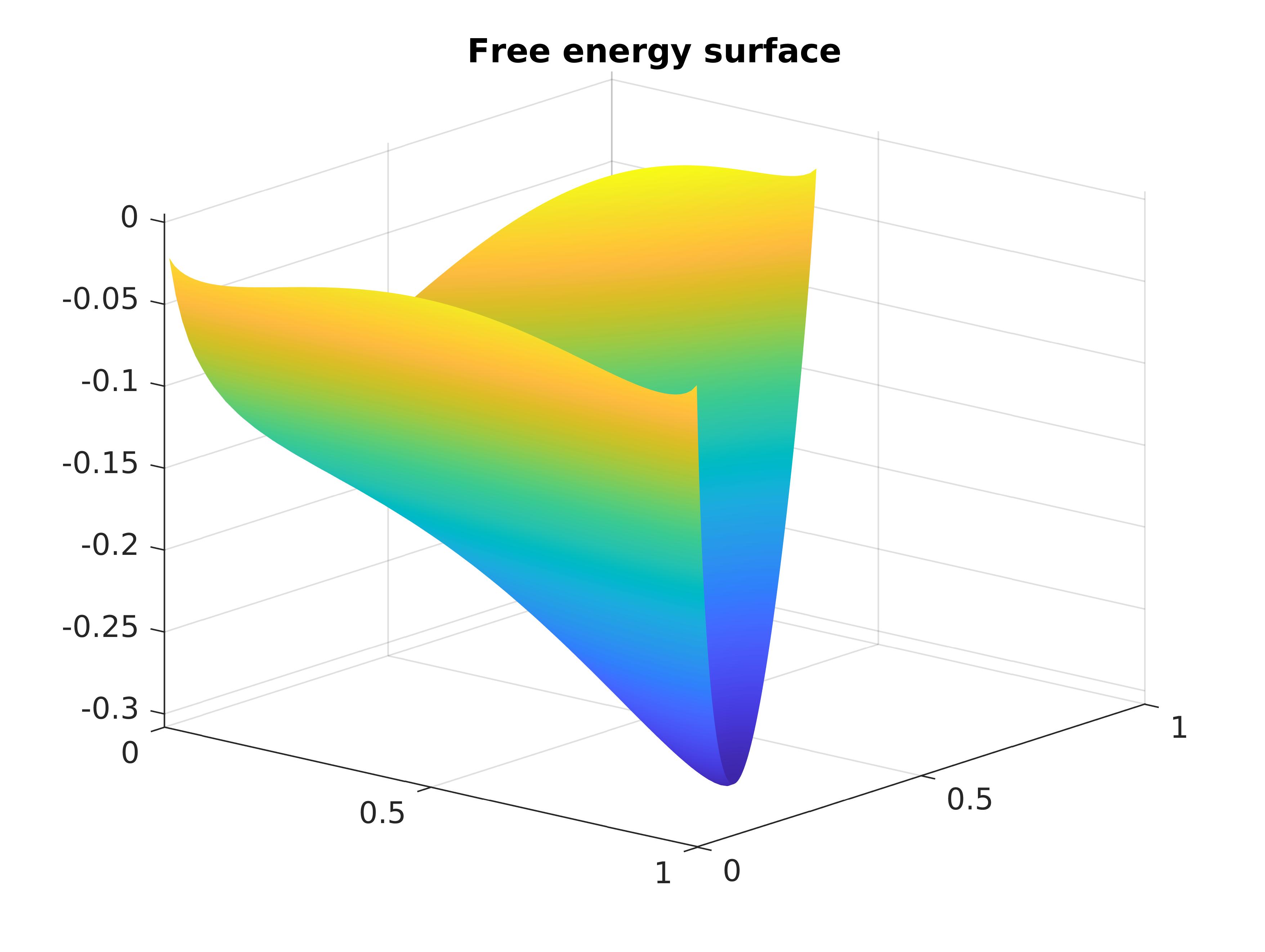}
        \includegraphics[width = 0.25\textwidth,trim={9cm 0 0 7cm},clip]{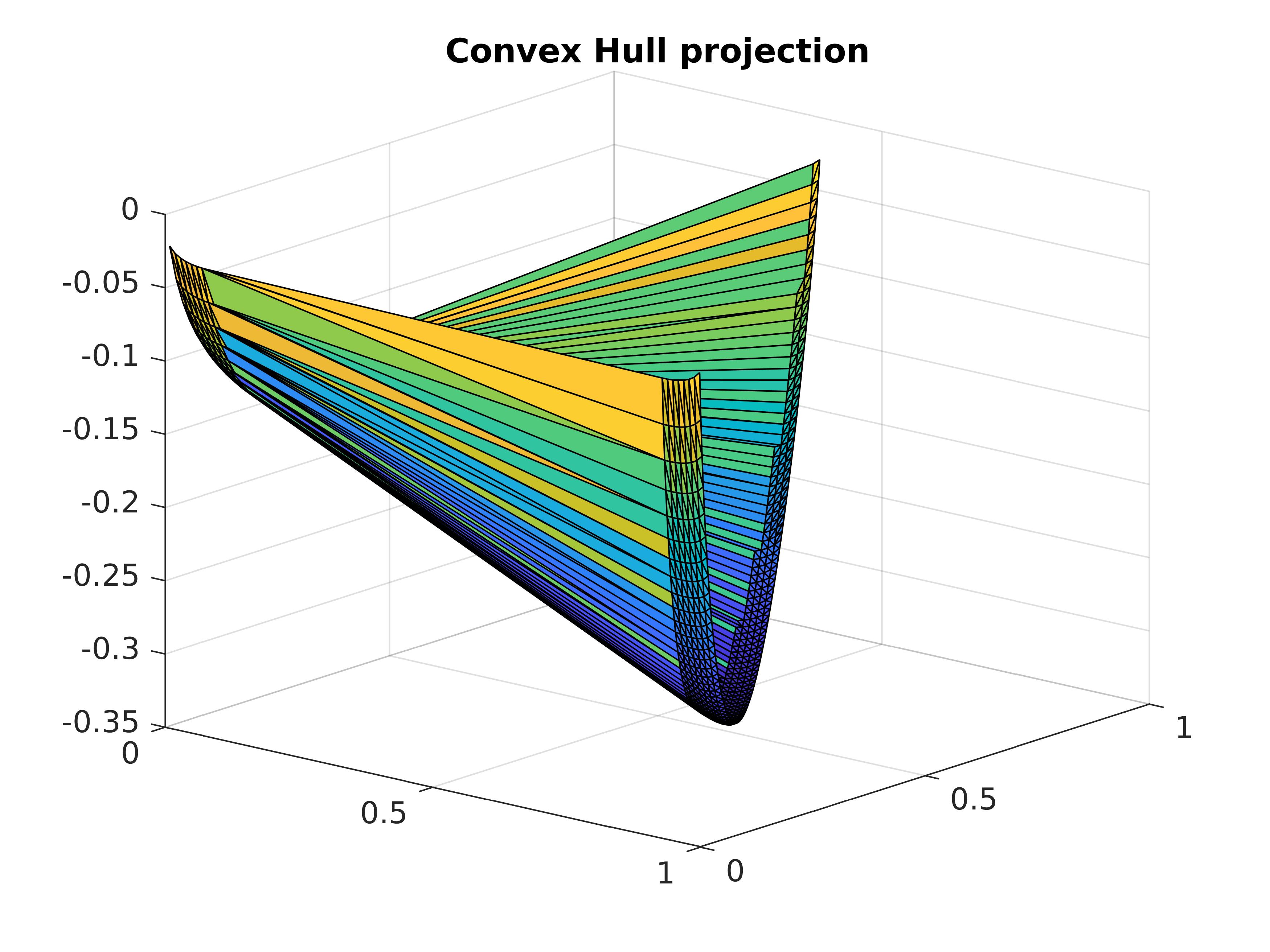}
        \includegraphics[width = 0.2\textwidth,trim={0cm 0 0 6.6cm},clip]{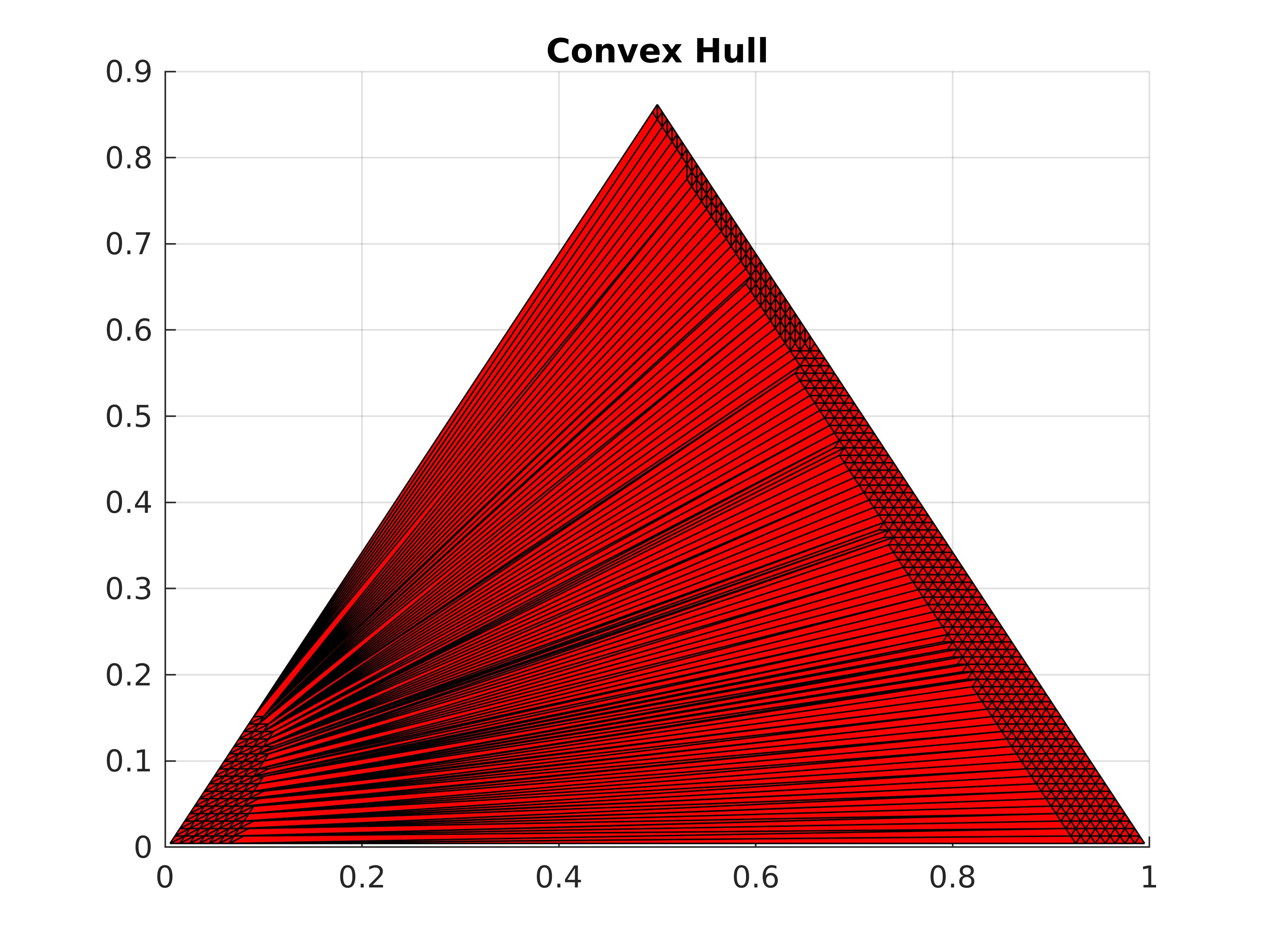}
        \includegraphics[width = 0.22\textwidth]{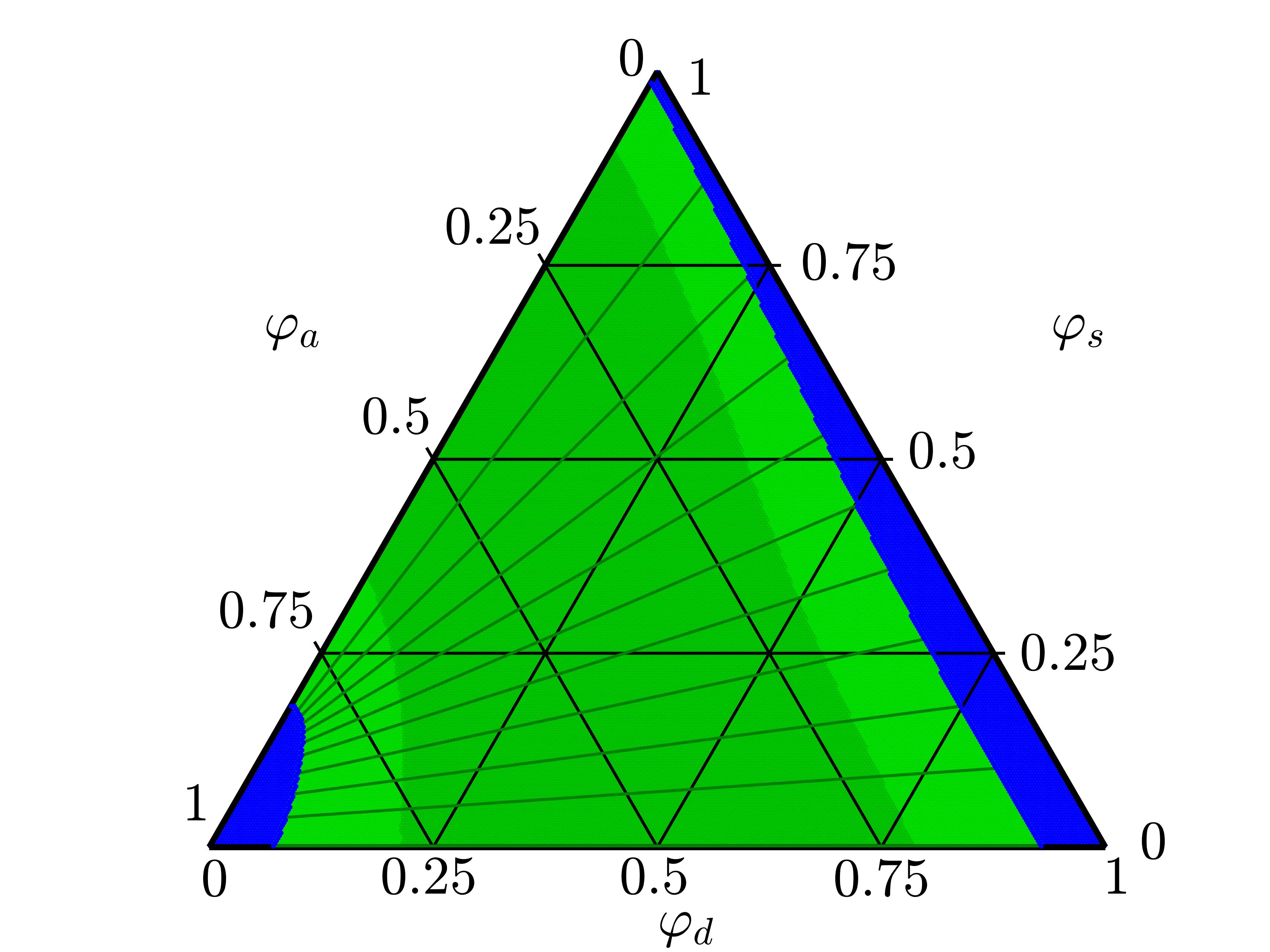} \\
        \includegraphics[width = 0.25\textwidth,trim={9cm 0 0 7cm},clip]{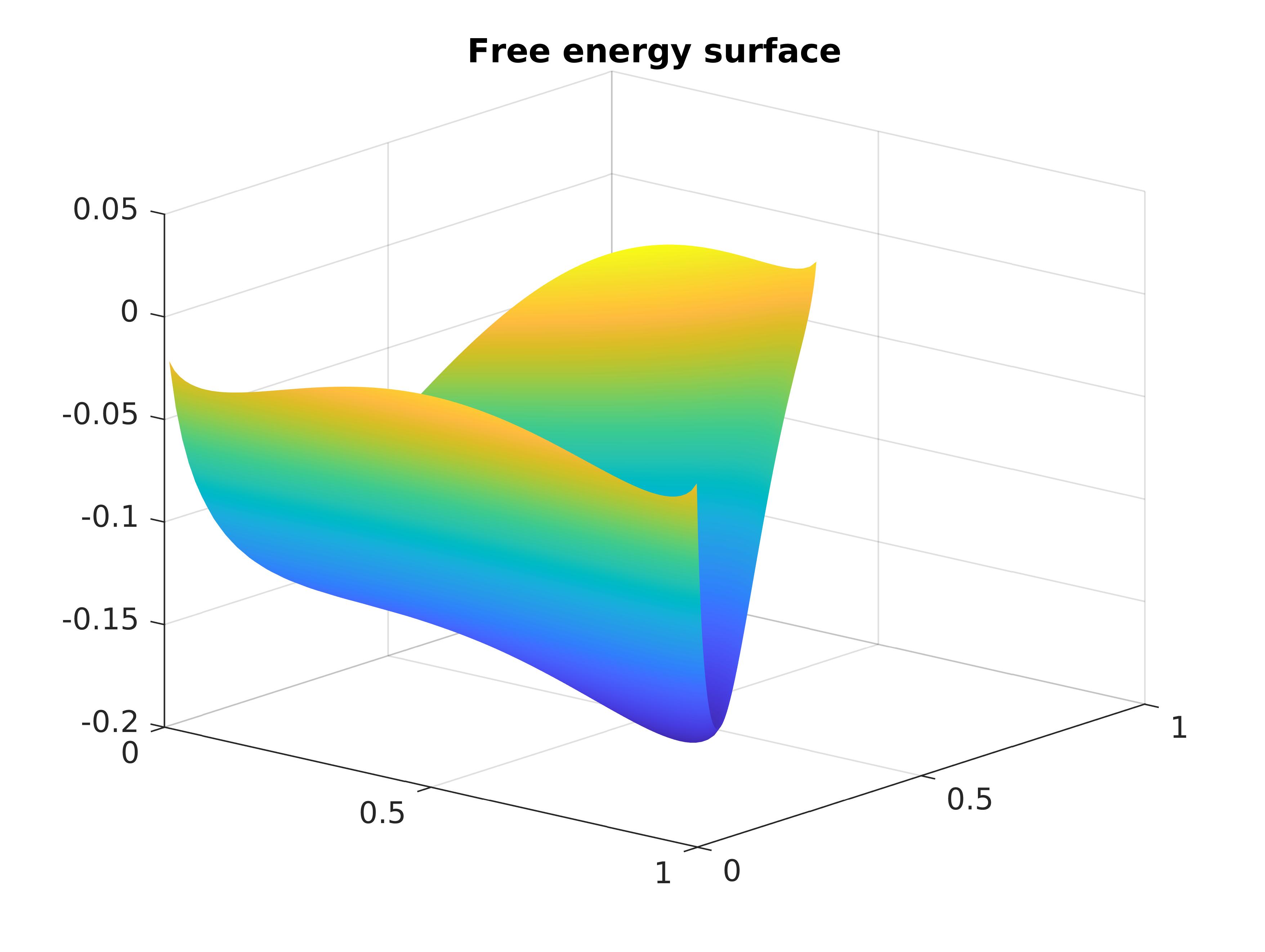}
        \includegraphics[width = 0.25\textwidth,trim={9cm 0 0 7cm},clip]{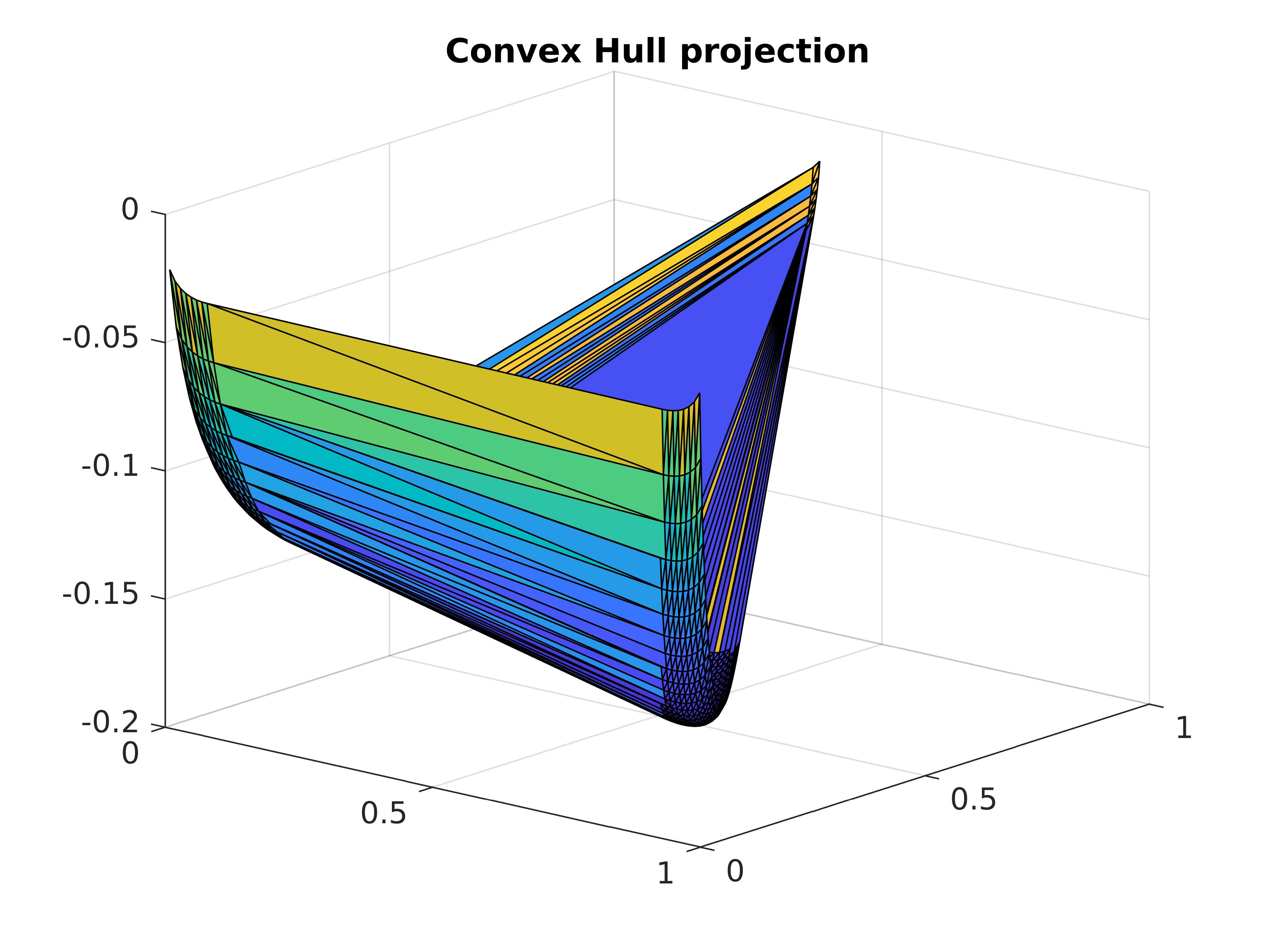}
        \includegraphics[width = 0.2\textwidth,trim={0cm 0 0 6.6cm},clip]{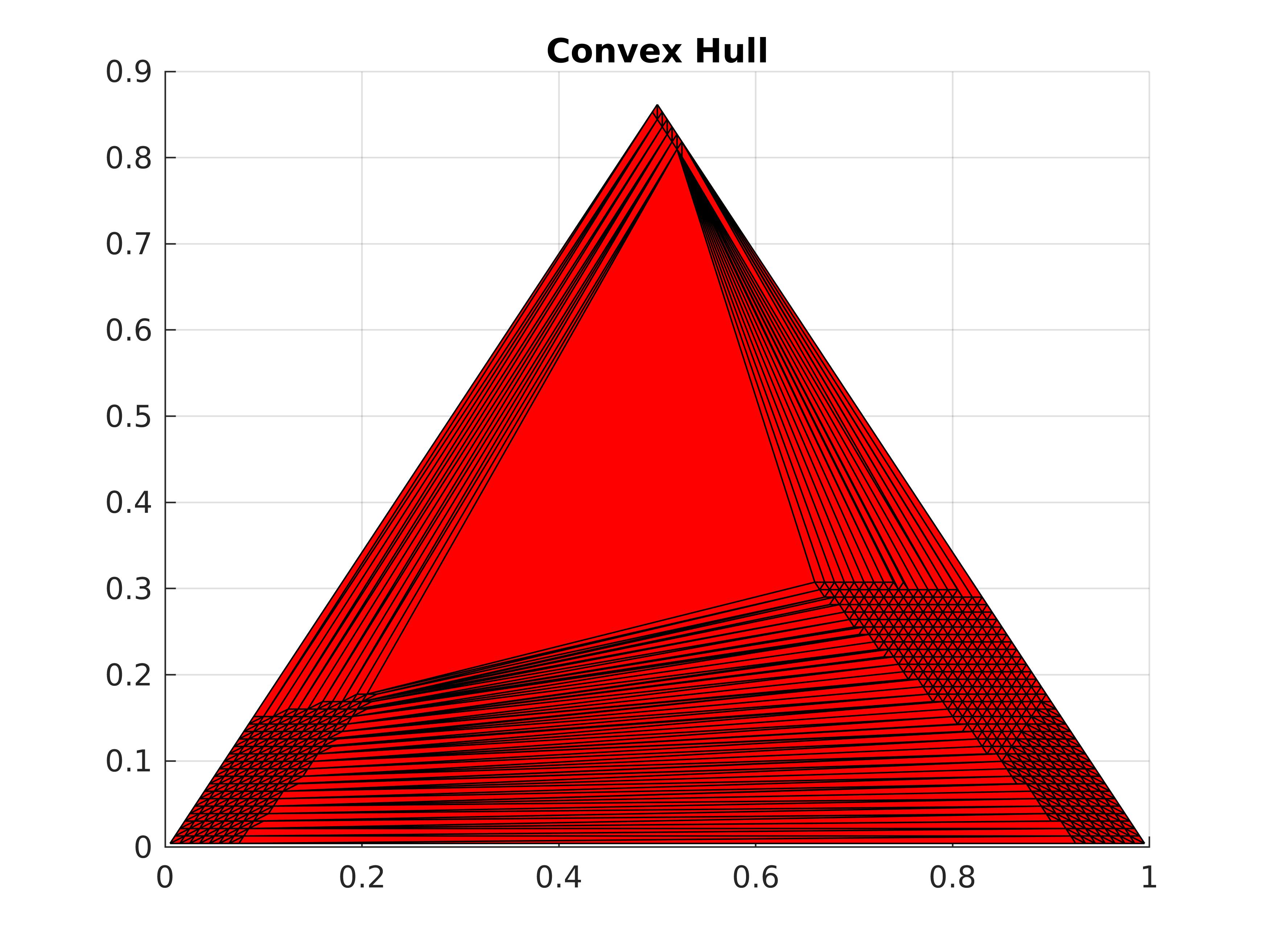}
        \includegraphics[width = 0.22\textwidth]{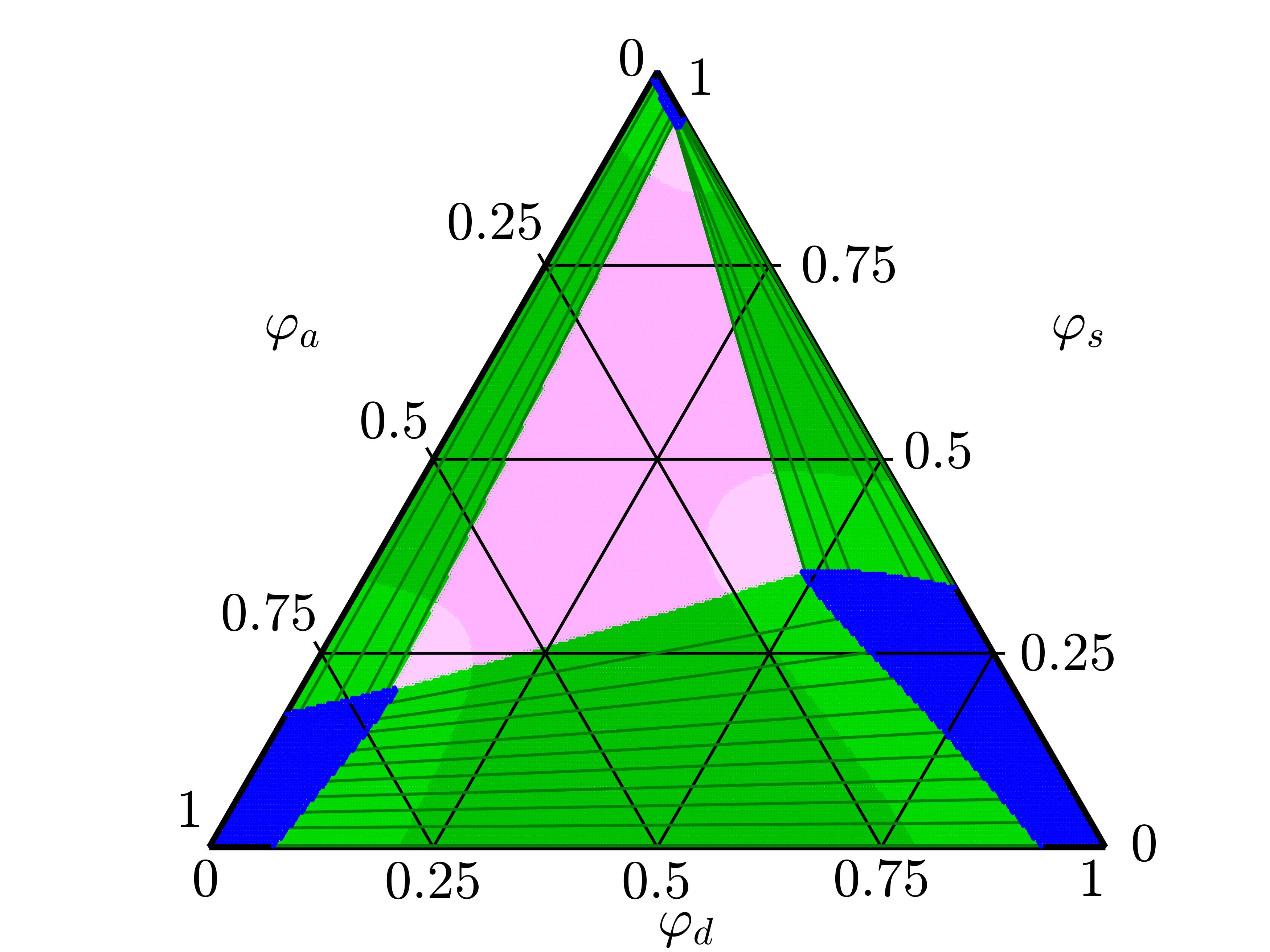} \\

    \caption{Phase diagram construction: (From left to right column) free energy surface, lower convex hull of the free energy surface, lower convex hull projection onto the composition space, phase diagram with (blue) stable one-phase regions, (green) two-phase equilibrium regions, (pink) three-phase equilibrium regions. (From top to bottom rows) Fully miscible blend ($\chi_{12}=-0.4$, $\chi_{13}=0.19$, $\chi_{23}=0$), blend with two-phase equilibrium region featuring a critical point, classical textbook example ($\chi_{12}=1.2$, $\chi_{13}=0.8$, $\chi_{23}=0$), blend with two-phase equilibrium region and no critical point ($\chi_{12}=0.6$, $\chi_{13}=0.52$, $\chi_{23}=1.57$), blend with three-phase equilibrium region ($\chi_{12}=0.6$, $\chi_{13}=1.25$, $\chi_{23}=1.6$). All phase diagrams are generated with 300 grid points in each direction for a small molecule - small molecule - solvent mixture ($N_1=N_2=5$, $N_3=1$).} 
    \label{fig:PDconst2}
\end{figure}

The phase diagram construction is based on the convex hull approach (see Figure~\ref{fig:PDconst2})~\cite{vaddi2023construction,voskov2015ternapi,mao2019phase,gottl_convex_2023}.
First, the volume fraction space is discretized using a regular triangular grid with the same number of grid points in all three compositional directions. 
Second, the free energy surface is evaluated at the discrete grid points using the Flory-Huggins equation (Eq.~\ref{eq:FloryHuggins}). 
Third, a convex hull of the free energy surface is determined. In this context, the lower part of the convex hull that coincides with the free energy landscape corresponds to the set of stable points of the composition space (one-phase equilibrium region). 
Fourth, the remainder of the lower convex hull is further processed to identify two-phase and three-phase equilibria regions. 
The vertices of this part of the convex hull are projected onto the composition space. 
This projection comprises two types of vertices: triangles with two elongated edges and triangles with three elongated edges that are larger than the initial grid spacing. 
The grid points of the last two types of triangles correspond to two-phase equilibria and three-phase equilibria regions, respectively. 
Note that for two-phase equilibria regions, the elongated segments of the convex hull triangles are good approximations of the tie lines if the grid is sufficiently fine. 

Four examples of phase diagrams with varying complexity are shown in the four rows of Figure~\ref{fig:PDconst2}.
From left to right, the free energy surface, the lower part of the convex hull, its projection to the composition space, and the corresponding phase diagrams are shown. The first phase diagram is typical of a fully miscible system, with the free energy landscape matching the lower part of the convex hull, resulting in only one stable region covering the whole phase diagram (marked with blue color). The second phase diagram in the figure is the typical textbook example of an amorphous ternary system with one two-phase region related to one pair of immiscible components. 
The free energy landscape contains a negative curvature region visible on the left panel in the second row of Figure~\ref{fig:PDconst2}. In the second panel, it can be observed that the lower convex hull wraps around this concave region with straight segments, actually connecting the composition of the two stable phases. The corresponding elongated vertices are more apparent from the third panel, which shows the projection of the lower part of the convex hull. The comparison between the third and the fourth panels shows how the two-phase region and the tie-lines correspond to the projection of the elongated vertices. 
The third example shows how the two miscibility gaps related to two different pairs of materials (concave regions visible on both the front and back-left boundary of the free energy surface on the first panel) can merge into a single miscibility gap. Finally, the last example contains one-two-phase and three-phase regions. In this case, looking at the free energy landscape alone is insufficient to suggest the phase diagram with the present complexity.

The accuracy of the phase diagram determination increases with the number of grid points. Despite this, several numerical artifacts might occur in rare or extreme cases. For example, for highly immiscible mixtures, the binodal lines may come so close to the domain boundaries that the neighboring stable regions are not displayed in the phase diagram. Moreover, the determination of the different regions is slightly sensitive to the criteria chosen to sort the triangles into small ones and large ones with two or three elongated edges. This is generally not problematic, except for phase diagrams associated with $\chi$ values close to the critical values $\chi^c$, which might result in low-curvature domains of the free energy surface. In such cases, calculating the convex hull leads to regions where the identification of the triangles is ill-defined and where grid refinement only partly helps. Very rare pathological consequences typically include a significant imprecision in determining the two-phase regions around the critical point or incorrect detection of three-phase equilibria regions within two-phase equilibria regions. Overall, approximately 0.5\% of the 81,000 phase diagrams generated in the present work were judged incorrect and could not be considered in the analysis.

%%%%%%%%%%%%%%%%%%%%%%%%%
\subsection{Data generation}
Three libraries of ternary phase diagrams are generated. They stand for three classes of material systems of interest in the field of printed electronics: polymer - small molecule - solvent (P-SM-S), small molecule - small molecule - solvent (SM-SM-S), and small molecule - solvent - solvent (SM-S-S) blends. The molecules differ by their molar size. Molar sizes of $N=1$, $N=5$, and $N=245$ are chosen as representative values for a solvent, a 'small' organic molecule material (larger than the solvent), and a polymer, respectively. 
We use material classes polymer, small molecule, and solvent to define different material systems based on the combination of molar sizes. However, the analysis is generic for a given combination of molar sizes. 
For example, $N_1=N_2=N_3=1$ can refer to a polymer-polymer-polymer system, a small molecule-small molecule-small molecule system, or, in principle, any other blend with three materials of similar molar size. Nevertheless, we will refer to it as a solvent-solvent-solvent system in the following.
The temperature is set to $300 K$ and the molar volume of the solvent to $v_{0}=10^{-4}$ $m^{3}/mol$, which is a representative value for classical solvents (e.g., toluene). The volume fraction space is discretized for all phase diagrams with 300 points along each volume fraction direction. For each library, the three interaction parameters are varied between $\chi_{ij}^{min}=\chi_{ij}^{c}-2\chi_{ij}^{c}$ and $\chi_{ij}^{max}=\chi_{ij}^{c}+2\chi_{ij}^{c}$, whereby $\chi_{ij}^c$ is the critical interaction parameter defined previously. 
Normalizing the space using the critical value of the interaction parameters provides a unified way of comparing the three material systems. 
In Figures~\ref{fig:allAboveAllBelow}-\ref{fig:sensitivity} below, the results are presented in the normalized parameter space using the variables $\bar{\chi}_{ij}=\chi_{ij}/\chi_{ij}^{c}$ ($\bar{\chi}_{ij}=1$ thus corresponding to the critical value). 

As discussed in detail in the results section below, the type of phase diagram strongly varies when the interaction parameters are close to the critical values. To capture these variations, logarithmic sampling is performed with a higher sampling density close to the three planes corresponding to the critical interaction parameter values in the 3D parameter space.
In total, 27,000 phase diagrams are generated for each material system considered (P - SM - S, SM - SM - S and SM - S - S). 
The parameters of the material systems are summarized in Table~\ref{tab:ParamSpace1}. 

Beyond the three libraries of phase diagrams for P-SM-S, SM-SM-S, and SM-S-S material systems, which consist of all phase diagrams from the systematic search over the range of interaction parameters summarized in the table below, additional sets of phase diagrams have been generated for other systems (P-S-S and S-S-S). Nevertheless, these additional phase diagram sets are limited to small pre-selected regions of the interaction parameter spaces.

\begin{table}[H]

\begin{center}
\begin{tabular}{
|p{0.3\textwidth}
|p{0.15\textwidth}
|p{0.15\textwidth}
|p{0.15\textwidth}|}
\hline 
\diagbox{\makecell[l]{Material\\ parameters}}{\makecell[l]{Material\\ system}} & \textbf{P-SM-S} & \textbf{SM-SM-S}& \textbf{SM-S-S}  \\
\hline
$N_{1}$ & $245$ & $5$ & $5$\\
$N_{2}$ & $5$ & $5$ & $1$\\
$N_{3}$ & $1$ & $1$ & $1$\\ 
$\chi^{c}_{12}$   &$0.131$ & $0.4$ & $1.047$\\
$\chi^{c}_{13} $&$0.566$ & $1.047$ & $1.047$\\ 
$\chi^{c}_{23}$  & $1.047$ & $1.047$ & $2$\\ 

\hline 
\end{tabular}
\caption{Material parameters used in this study for three material systems: P-SM-S, SM-SM-S, and SM-S-S (P-polymer, SM-small molecule, S-solvent). The molar size and the corresponding binary critical interaction parameters used to define the design space ($\chi_{ij}^{min}=\chi_{ij}^{c}-2\chi_{ij}^{c}$ and $\chi_{ij}^{max}=\chi_{ij}^{c}+2\chi_{ij}^{c}$) are listed in the table.}
\label{tab:ParamSpace1}
\end{center}
\end{table}

%%%%%%%%%%%%%%%%%%%%%%%%%
\section{Results and discussion}

\subsection{Identification and classification of phase diagram types}
\begin{figure}[H]
    \centering
\includegraphics[width=\textwidth, trim=1 425 1 4, clip]{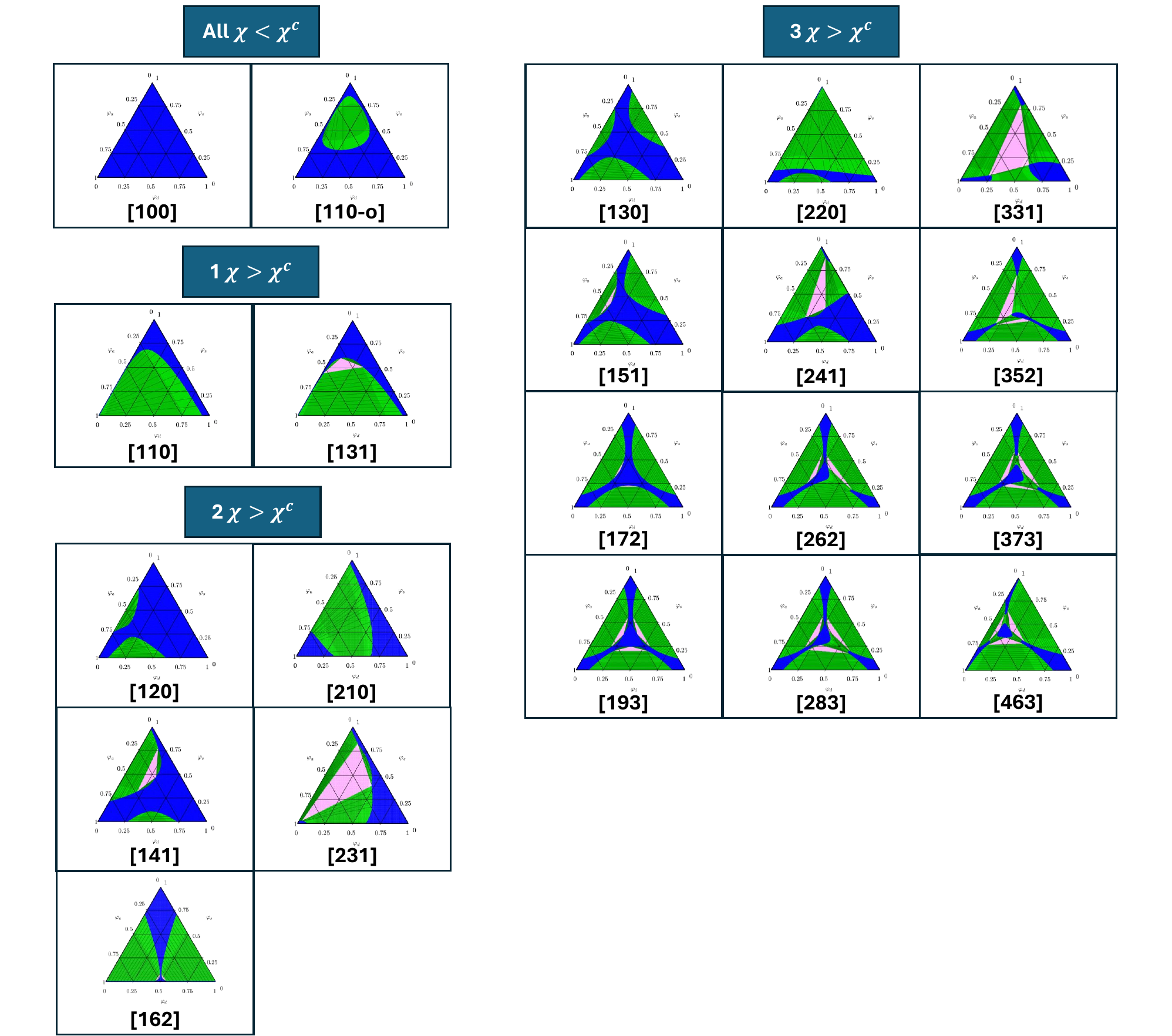}  
    \caption{Classification of phase diagram types identified in the library, depending on the number of immiscible material pairs (as marked in the panels). The corresponding three-digit key is reported below each type of phase diagram. The first digit stands for the number of one-phase regions, the second for the number of two-phase equilibrium regions, and the third one for the number of three-phase equilibrium regions. See SI-4 for the material properties of each phase diagram. }
    \label{fig:zoology}
\end{figure}

Figure~\ref{fig:zoology} provides an overview of the phase diagram types identified for the material systems investigated in this work. Each phase diagram depicts regions with a distinct number of equilibrium phases. The blue regions correspond to one-phase stable regions, the green regions to two-phase equilibrium regions (deep green for unstable and light green for metastable regions), and the pink regions to three-phase equilibrium regions (deep pink for unstable and light pink for metastable regions). 
For the two-phase equilibrium regions, the tie lines connecting the compositions of equilibrium phases are also shown. 
Each phase diagram in Figure~\ref{fig:zoology} is identified by a three-digit key, where each digit represents the number of distinct regions with one-phase, two-phase, and three-phases. 
For example, the type of phase diagram marked with [331]-key, which is typical of a system with three pairs of immiscible components and is presented at the top right corner of Figure~\ref{fig:zoology}, contains:
(i)~3 distinct regions marked blue that correspond to three one-phase regions at the corners of the phase diagram,
(ii)~3 distinct two-phase regions marked green that are located along each side of the phase diagram and
(iii)~1 central three-phase region color-coded pink.

Figure~\ref{fig:zoology} summarizes the 21 phase diagram types identified within the generated libraries using a connected component-based method that determines the number of one-phase, two-phase, and three-phase equilibrium regions in the phase diagrams. 
The method first identifies distinct regions within the phase diagram using the connected components algorithm~\cite{cormen2022introduction}. 
Then, each region is labeled with the number of equilibrium phases.
As a result, each phase diagram is assigned a three-digit number representing the number of one-, two-, and three-phase regions (connected components). This three-digit number is used for phase diagram classification. The method successfully classified most phase diagrams, except phase diagrams with numerical artifacts, explained at the beginning of the Method section, where semi-manual classification was used.

Figure~\ref{fig:zoology} is also the first visualization of the existence rules for the ternary phase diagrams, sorted hierarchically. Thereby, the phase diagrams have been classified by (1)~the number of immiscible materials pairs ($\chi_{ij}>\chi_{ij}^c$ for materials $i$ and $j$, see Eq.~\ref{eq:CriticalInteractionParameter}), which is the first-order existence rule, (2)~the number of distinct miscibility gaps found in the phase diagrams, and (3)~the number of distinct three-phase regions. Below, the mapping between the design space and the types of phase diagrams is described in more detail. 

Starting with the simplest case -- with three miscible pairs of components (all $\chi<\chi^c$), the most common phase diagram is the one with zero miscibility gap. A single one-phase (blue) region covering the composition space is observed. The three-digit key for this type is [100] - as marked in the first panel of Figure~\ref{fig:zoology} with the phase diagram being fully blue. This phase diagram is very common: the fact that the three material pairs are miscible naturally promotes the miscibility of the ternary mixture. However, even with all $\chi<\chi^c$, phase diagrams with one closed-loop two-phase region and two critical points~\cite{rabeony1994closed} can be found, even if they are much less common (see next section). This phase diagram is denoted as [110-o], whereby "-o" stands for the closed-loop shape of the miscibility gap. 

When one pair of components is immiscible (one $\chi>\chi^c$), a single miscibility gap is obtained in the phase diagram. The miscibility gap features one critical point (where the binodal and spinodal curves intersect). Most often, this is a two-phase equilibrium region. This type of phase diagram, the "classic textbook example" for the ternary phase diagram of polymer solutions, is denoted with the key [110]. In less common situations, even though only one pair of immiscible materials exists, the miscibility gap can feature a three-phase region (type [131]).

When two pairs of components are immiscible (two $\chi>\chi^c$), two miscibility gaps (with two critical points) might be present. Most often, these miscibility gaps correspond to two-phase regions. This type of phase diagram is denoted with a [120] key. Here again, a three-phase region might be present in one of the miscibility gaps (type [141]). Two three-phase regions might also be present, one in each miscibility gap (type [162]). However, when the interaction parameters are sufficiently large, both miscibility gaps merge. Note that the critical points disappear as a result of the merging. If the miscibility gap does not feature any three-phase region, this results in a [210] phase diagram type. This is the most frequent phase diagram for two pairs of immiscible components (see next section). In rare cases, one three-phase region might be present (type [231]). Even if we could not report them from our current library, we expect phase diagrams with one merged miscibility gap and two three-phase regions to be possible (type [252]). 

Finally, when all three pairs of components are immiscible (3 $\chi>\chi^c$) and the interaction parameters are sufficiently low, three miscibility gaps are found at the edges of the phase diagram. Each miscibility gap may feature a three-phase region, which results in types [130] (zero 3-phase region), [151] (one 3-phase region), [172] (two 3-phase regions) or [193] (three 3-phase regions). With typically one of the interaction parameters becoming larger, two of the miscibility gaps can merge, resulting in phase diagram types [220] (zero 3-phase region), [241] (one 3-phase region), [262] (two 3-phase regions), or [283] (three 3-phase regions). With even larger interaction parameters, all three miscibility gaps merge. A single three-phase region frequently lies in the middle of the miscibility gap (type [331]). Nevertheless, phase diagram configurations with two (type [352]) or three (types [373] and [463]) three-phase regions can be encountered. In the latter case, the three miscibility gaps are slightly in contact, and a miscible region can be found in the middle of the miscibility gap. 

Figure~\ref{fig:zoology} presents 21 possible types of phase diagrams with varying numbers of miscibility gaps (from 0 to 3) and three-phase regions (up to 3). Altogether, to our knowledge, this is the most comprehensive and extensive list of phase diagrams for ternary amorphous blends reported so far, even though the free energy function considered in this work is remarkably simple. A similar library of phase diagram types has been reported previously for fluid ternary systems~\cite{ryll2012convex}, with most phase diagrams consistent with this work. The phase diagram types [193], [130], and [463] were also simulated as well by Huang and coworkers \cite{huang_phase_1995}. However, these previous works only presented and/or investigated a subset of the phase diagrams reported in the current work. Despite our systematic and extensive screening, we suspect some other (very rare) types of phase diagrams to be possible, which might exist in very narrow zones of the parameter space that were not sampled in our libraries.

%%%%%%%%%%%%%%%%%%%%%%%%%%%%%%%%%%%%%%%%%%%%%%%%%%%%%%%%%%%%%%%%%%%%%%%%%%%%%%%%
\subsection{Octant-based existence rules are universal over 
multiple material systems}
The observations made so far can be considered as qualitative existence rules. 
These rules are based on the number of immiscible pairs of components and can be mapped to the octants of the design space. In the following, the design space is split into eight octants using the three planes at the critical parameter values as the octant boundaries. The design space is presented in the normalized interaction parameter space $(\bar{\chi}_{12},\bar{\chi}_{13},\bar{\chi}_{23})$, and its center is located at $(1,1,1)$ (see data generation subsection). The octant-based visualization in the normalized interaction parameter space $(\bar{\chi}_{12},\bar{\chi}_{13},\bar{\chi}_{23})$ will be used in the following subsections.

\begin{figure}[H]
    \includegraphics[width=0.95\textwidth]{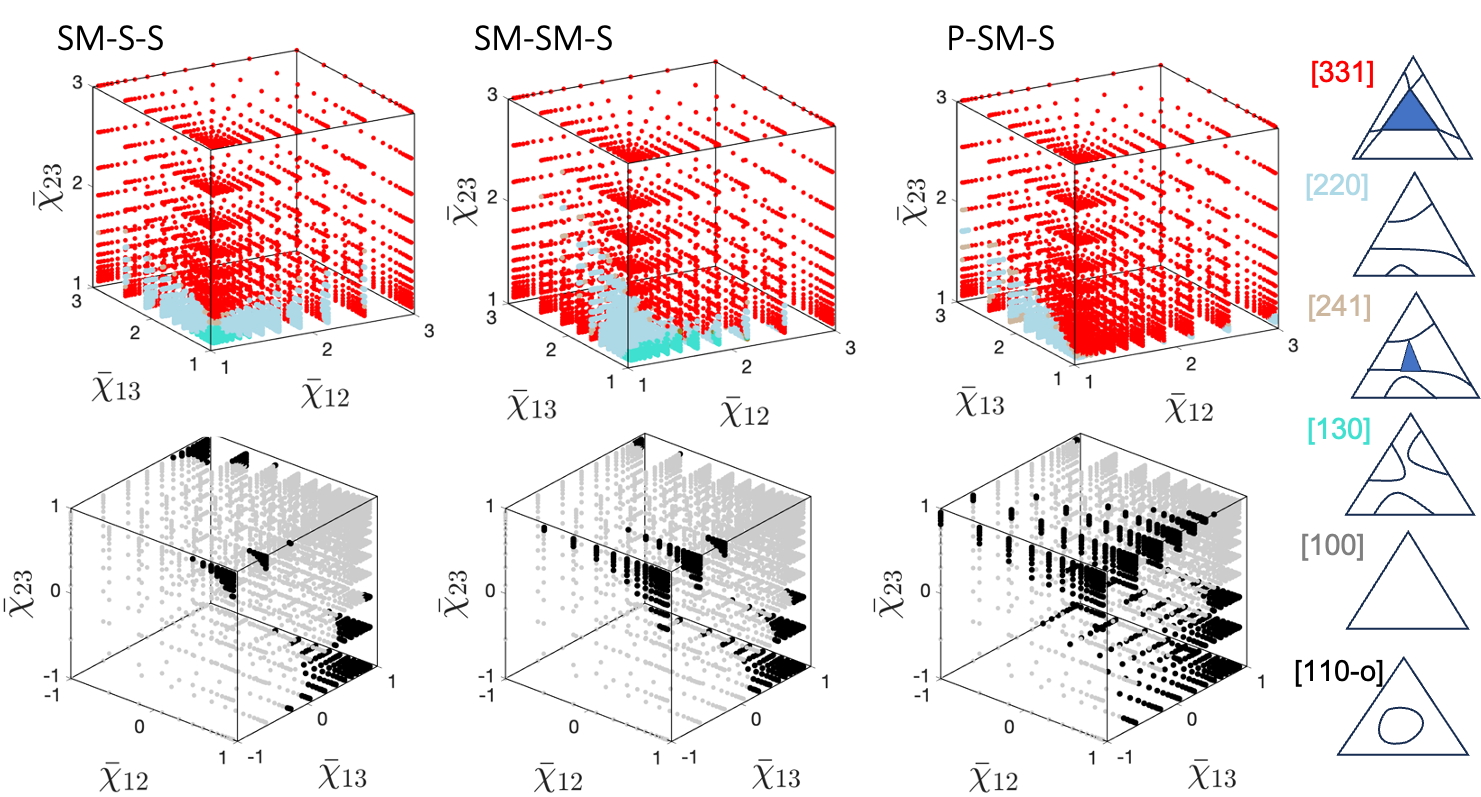}
    \caption{Two octants of design space: three $\chi>\chi^{cr}$ (top), no $\chi>\chi^{cr}$ (bottom) mapped to the six types of phase diagrams. The types are schematically shown in the right column, and the three-digit keys are provided. The results are provided for three materials systems (SM-S-S, SM-SM-S, and P-SM-S) in the first three columns of the figure.}
    \label{fig:allAboveAllBelow}
\end{figure}

With the above normalization and visualization defined, the first-order existence rules are also called octant-based existence rules because of the mapping between them and the octants of the design space. As detailed below, the octant-based existence rules are found to be universal across the three materials systems analyzed in this work. This subsection provides the detailed mapping between octants of the design space and types of phase diagrams.

Starting with the octant where three pairs of components are immiscible, $\chi_{ij} > \chi_{ij}^c$, the phase diagram type [331] is the dominant (most likely) type, regardless of the material system: for the three systems, SM-S-S, SM-SM-S, and P-SM-S, as depicted in Figure~\ref{fig:allAboveAllBelow} top row, the points color-coded red dominate most of the octant space.
The second most common phase diagram is [220], and it is most commonly found in the region where at least two interaction parameters are close to their critical values. The least common phase diagrams are of type [130], [151], [241]. A more detailed discussion of these rare phase diagrams and their location in the design space (Figure~\ref{fig:leastCommon}) is provided later in this section. 

Moving to the octant where three pairs of components are miscible, $\chi_{ij} < \chi_{ij}^c$, the phase diagram type [100] is the dominant type (grey points in Figure~\ref{fig:allAboveAllBelow} - bottom row), regardless of the material system. The second most common phase diagram is of type [110] with closed-loop immiscibility (black points in Figure~\ref{fig:allAboveAllBelow} - bottom row). 
For SM-SM-S and SM-S-S systems, this type of phase diagram is found when one interaction parameter is negative, and two others are close to their critical values. Moreover, the region of existence for this type of phase diagram becomes larger for lower values of the negative interaction parameters. For the P-SM-S system, a similar behaviour is observed, however this type of phase diagram is observed even for positive values of $\chi_{13}$. The division of the design space into regions of type [100] and [110-o] is consistent with recently published work~\cite{zhang2024phase} that discusses the effect of co-nonsolvency in polymer - solvent - cosolvent mixtures where all components are pair-wise miscible. 
It has been shown by Zhang~\cite{zhang2024phase} that the boundaries separating type [100] and [110-o] are curved, which is confirmed by our findings. Additionally, the region of the design space with the closed-loop immiscibility has been found to become larger as the molar size increases, which is consistent with our analysis: for the P-SM-S system, a significantly larger region of the screened domain is marked with black points corresponding to this type of phase diagram (Figure~\ref{fig:allAboveAllBelow} - bottom panels).

\begin{figure}
    \includegraphics[width=1\textwidth]{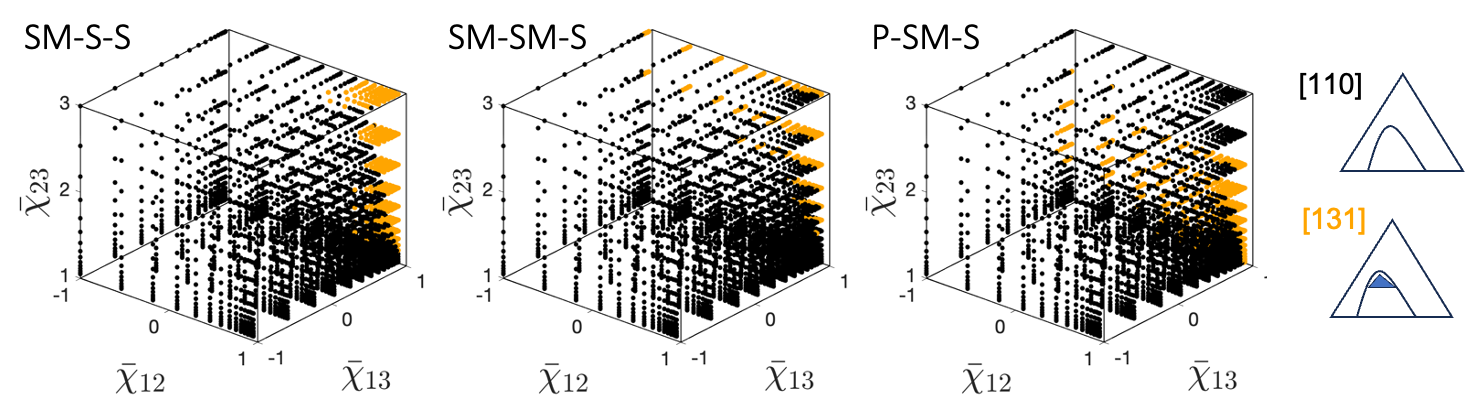}
Figure 4: Representative octant of design space with one    \caption{Representative octant of design space with one $\chi>\chi^{cr}$. Three panels correspond to three material systems: SM-S-S, SM-SM-S and P-SM-S. Note that the distribution within the design space is reported for two-phase diagrams [110] and [131], with the latter containing one three-phase region. The legend to the right provides the schematics. The three-digit keys are colored to match the points in the design.}
    \label{fig:oneAbove}
\end{figure}

When one interaction parameter is above a critical value (one $\chi_{ij} > \chi_{ij}^c$), the 'classical textbook example' phase diagram [110] becomes the dominant type. Note that in the design space, three octants correspond to the criterion 'one $\chi_{ij} > \chi_{ij}^c$'. Since the observations are the same for the three octants, only one octant is shown in Figure~\ref{fig:oneAbove} for each material system. All other octants are included in the SI-1. This phase diagram with one miscibility gap is marked with black points and is undoubtedly the most prevalent and significant phase diagram for these three octants. The second phase diagram type is [131], which contains one three-phase region inside the miscibility gap - see orange points in Figure~\ref{fig:oneAbove}. The type [131] phase diagram is found only when the two miscible material pairs are characterized by an interaction parameter slightly below the critical value. Moreover, for both SM-S-S and SM-SM-S material systems, the existence region of this type becomes larger as the interaction parameter of the immiscible pair increases (Figure~\ref {fig:oneAbove} - left and center). For the P-SM-S system, however, it is noteworthy that a high interaction parameter of the immiscible pair leads to the emergence of the phase diagram of type [110]. In other words, [131] phase diagrams are not found for large interaction parameter values of the immiscible pair. This finding has significant implications for understanding and controlling phase behavior in these systems. When this transition occurs, the phase diagram of type [110] is additionally characterized by highly titled tie-lines -- see SI-2.

\begin{figure}
    \includegraphics[width=0.95\textwidth]{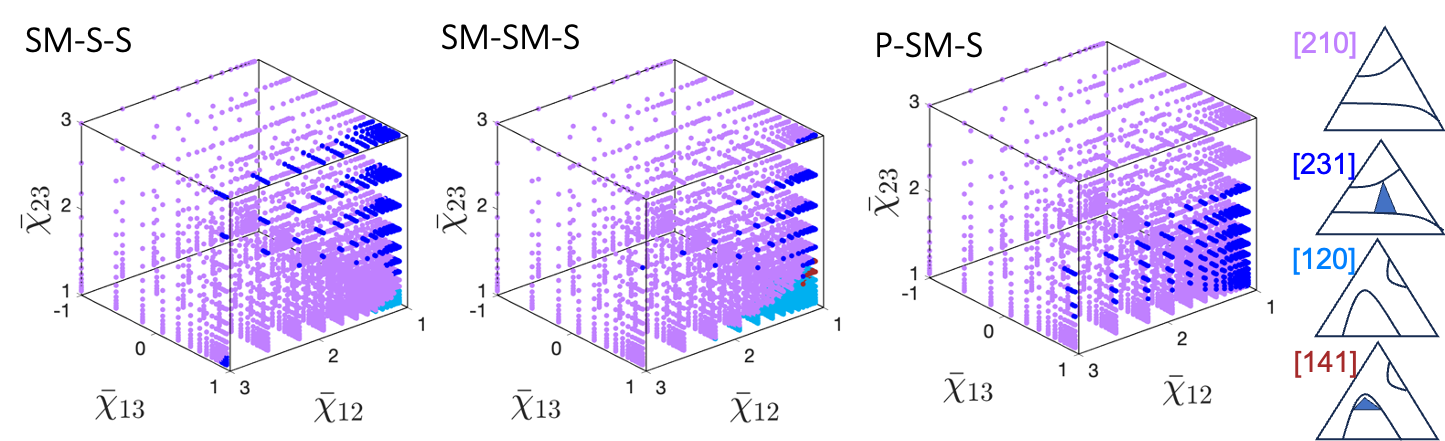}
    \caption{Representative octants of design space with two $\chi>\chi^{cr}$. Three panels correspond to three material systems: SM-S-S, SM-SM-S and P-SM-S. Note the distribution within the design for four types of phase diagrams: [210], [231], [120] and least common [141]. The legend to the right provides the schematics. The three-digit keys are colored to match the points in the design.}
    \label{fig:twoAbove}
\end{figure}

Moving to the last octants, when two pairs of the components are immiscible, the asymmetrical phase diagram of type [210] is the most common (two $\chi_{ij} > \chi_{ij}^c$). In Figure~\ref{fig:twoAbove}, these cases are marked with purple points. Here again, only one octant for each material system is shown in the figure. The visualization of the remaining octants can be found in SI-1. 
Two types of phase diagrams found in these octants - [120] and [231] - belong to the less common category. The type [120] can be seen when three interaction parameters are close to their critical values (right bottom rear corner of the design space in Figure~\ref{fig:twoAbove}). However, this type of phase diagram is found only for SM-S-S and SM-SM-S material systems. 

Type [231] is typical for cases where interaction parameters of miscible pairs are close to their critical values in the octant. 
Interestingly, the three material systems display slightly different distributions of the phase diagram types.
For SM-S-S system, as $\chi_{23}$ increases, this type becomes more prevalent in the plane ($\chi_{12}$,$\chi_{13}$). This trend is clear for the front part of the octants in Figure~\ref{fig:twoAbove} for type SM-S-S.
For SM-SM-S and P-SM-S systems, this trend is also reported but to a lesser degree. Moreover, as $\chi_{23}$ increases even further, the trend reverses. 
However, for $\bar{\chi}_{23}>2$, type [210] emerges and becomes dominant for high values of $\chi_{23}$. 
As a reminder, [231] type corresponds to the case when the two interaction parameters for the immiscible pairs increase in strength. 
The dominance of type [210] over [231] for significantly higher values is associated with the transition occurring when the tie-lines of two two-phase regions align in the comparable direction - see SI-2 for the details. 
We refer to the detailed reason behind this transition to another future study.
Finally, the type [141] is reported to be the least common phase diagram in this octant and can be identified in a very narrow range of interaction parameters.
This type occurs during the transition between type [120] and [210] and is only present for some octants reported in this work.

%%%%%%%%%%%%%%%%%
\subsection{ Less common types of phase diagrams for interaction parameters close to their critical values}
This paper reports three groups of phase diagram types: the most common, less common, and least common types. 
The most common types can be intuitively linked and understood through the number of immiscible pairs of components. 
In this subsection, we look closer at two other categories of phase diagrams, the less common and the least common types. 

\begin{figure}
    \includegraphics[width=0.95\textwidth]{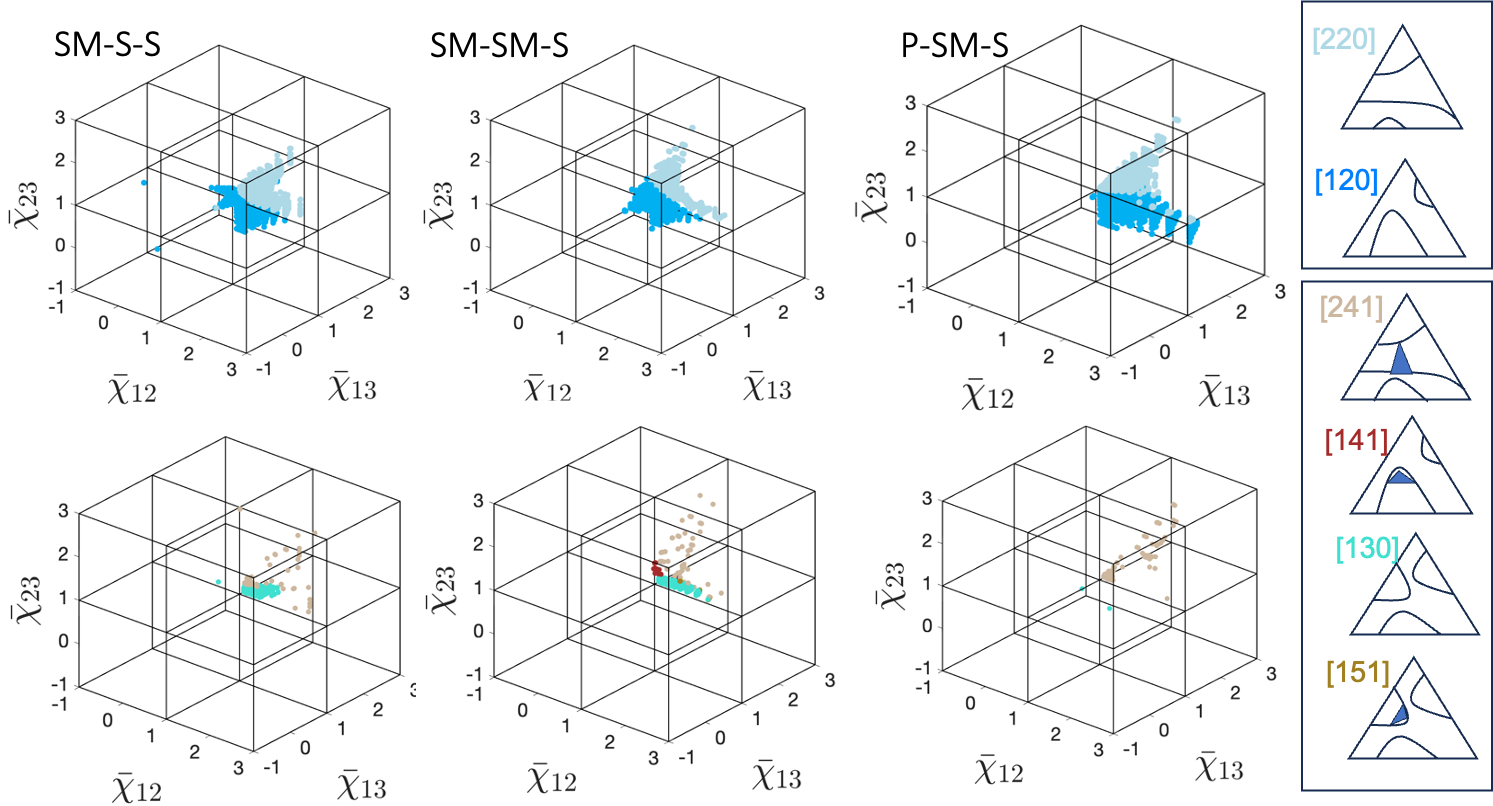}
    \caption{Distribution of less common (top) and infrequent (bottom) phase diagrams in design space of interaction parameters for three material systems. The legend to the right provides the schematics. The three-digit keys are colored to match the points in the design.}
    \label{fig:leastCommon}
\end{figure}

Two types of phase diagrams are considered less common: [220] and [120]. In Figure~\ref{fig:leastCommon}, their occurrences in the design space are marked with blue and light blue colors - see the legend in the figure. 
Phase diagrams [220] and [120] are identified close to the center of the normalized interaction parameters coordinate system for all three materials systems considered in this work. 
As already stated, the type [220] is identified only for cases when three interaction parameters are above critical values (upper right octant in the figure), with at least one interaction parameter being small. Type [120] exists within the octants with two pairs of immiscible components, and similar to [220], this phase diagram is identified for relatively low interactions between two immiscible components, yet above the critical values. 

The existence of both [220] and [120] types of phase diagrams respects the octant-based existence rules detailed above, although over smaller regions in the design space. This is also the case for the least common type of phase diagrams, which are present in a very narrow range of parameters and are mainly found in the octants with all three pairs of materials being immiscible. 
The type [130] phase diagram can be observed only when all three interaction parameters are slightly above the critical values.
This can be seen in Figure~\ref{fig:leastCommon} bottom panels, with points marked with light green points close to the (1,1,1) point. 
Moreover, this type of phase diagram with three distinct miscibility gaps and three critical points is less likely for the system with a higher molar size (type [130] is not reported here for the P-SM-S system). 
The three remaining types reported in this plot—[141], [241], and [151]—contain a three-phase region. 
These are also reported in the small region of the design space, where at least one interaction parameter is close to the critical value. 
Finally, only one phase diagram of type [463] has been identified among the 81,000 diagrams of the three libraries. This is evidence of an extremely narrow range of interaction parameters for which this phase diagram can be identified. 
Hence, although possible, observing these types in the design space of interaction parameters studied in this work is extremely unlikely. 

\subsection{Low miscibility depth in octants where two  $\chi>\chi^{cr}$}
\begin{figure}
    \includegraphics[width=1\textwidth]{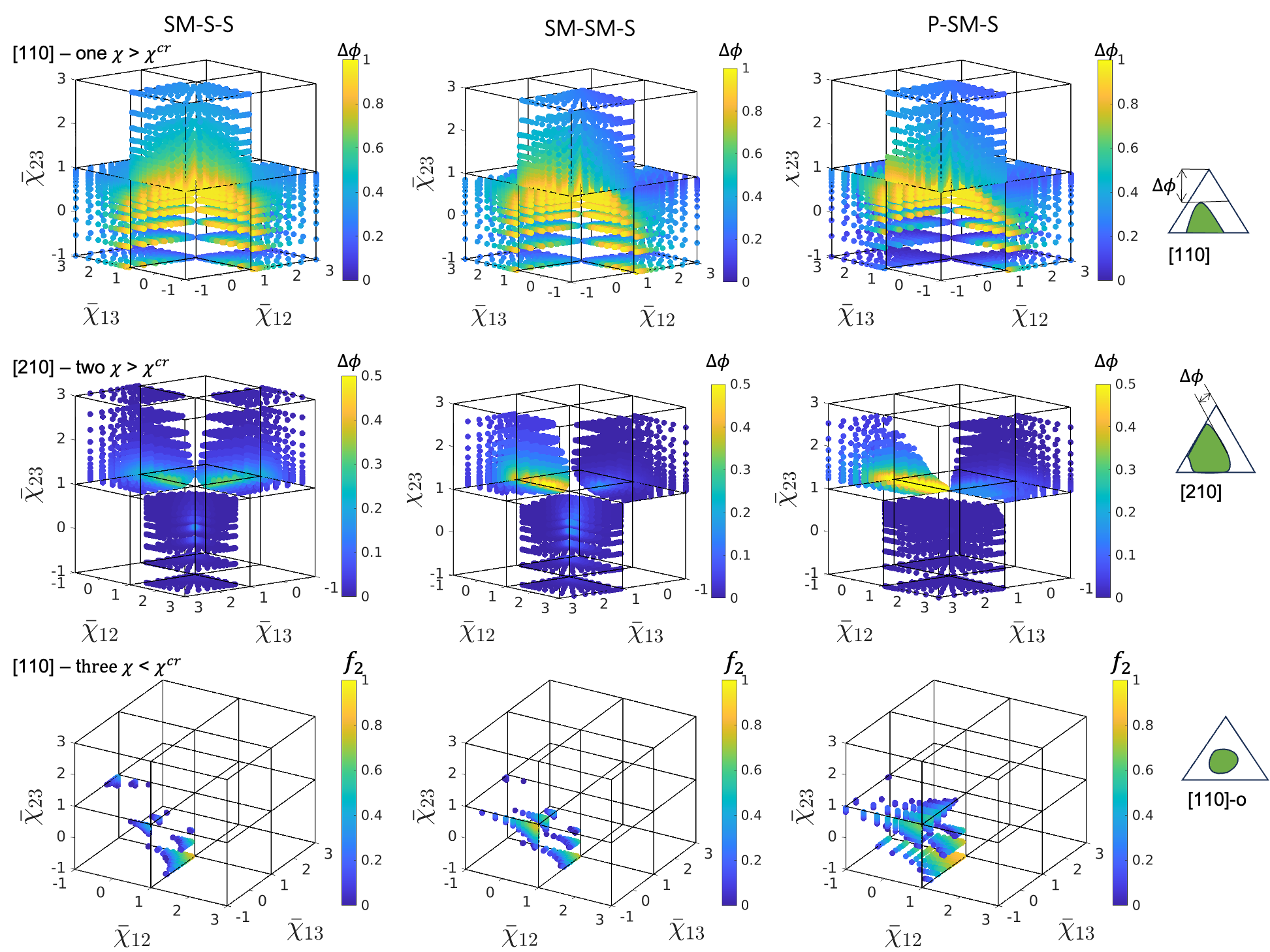}
    \caption{Sensitivity of phase diagram characteristics to changes in the interaction parameters for (top) [110] phase diagram with one miscibility gap, (middle) [210] phase diagram with two merged miscibility gaps, and (bottom) phase diagram with closed loop.  }
    \label{fig:sensitivity}
\end{figure}

Additional analysis of the phase diagram characteristics is reported for three selected types of phase diagrams, namely [210] [110], and [110-o] (with the closed loop miscibility gap). 
Figure~\ref{fig:sensitivity} depicts the miscibility depth $\Delta \phi$ for three types of phase diagram. Note that for types [110] and [210], results are depicted in three octants of the design space along the major direction.
Moreover, three permutations of one (or two) miscible pairs are possible for the ternary system. 
For type [110], the major compositional direction corresponds to the component miscible with two others (e.g., when $\chi_{12}>\chi_{12}^c$, the direction along $\phi_3$ is the major direction along with the miscibility gap is reported).
For type [210], the major compositional direction corresponds to the component that is immiscible with two others (e.g., when $\chi_{12}>\chi_{12}^c$, and $\chi_{13}>\chi_{13}^c$ direction along $\phi_1$ is the major direction along with the miscibility gap is reported). 
The insert on the right of the figure depicts the characteristics extracted, while SI-3 provides more details.
However, for three combinations of interaction parameters, the miscibility gap will be oriented differently, with the depth taken along the composition direction perpendicular to the edge corresponding to the immiscible binary. 
Similarly, for the material system with two immiscible pairs of components exhibiting a [210] type of phase diagram, the extracted depth will depend on the specific pairs.
The combined presentation of the miscibility depth allows one to visualize the sensitivity of the two-phase region to the choice of the interaction parameters. 
For type [110-o] with the closed loop, to present the sensitivity of the phase diagram to the changes in the parameters, the fraction of the phase diagram with the two-phase region is reported, $f_2$. The inset in the figure highlights the area of the loop that, when divided by the area of the phase diagram, gives $f_2$.

For three material systems and the type [110] phase diagram, the miscibility depth varies from a value as small as 0.0183 to 0.957 for the P-SM-S system, irrespective of the octant. 
Such a wide range of miscibility depth opens the opportunity to expand the manufacturing window. For example, for larger miscibility depth, significantly less solvent can be used to prepare the diluted solution for the casting.    
Interestingly, for the same material systems but type [210] phase diagrams, the miscibility depth ranges only from 0.000833 to 0.505. 
For two out of three octants, the range is even smaller. 
The figure also includes the dependence of the fraction of the two-phase closed loop region on the choice of interaction parameter for the phase diagram [110-o] (lower row). 
The fraction decreases s as the interaction parameter decreases and is higher for the P-SM-S system. 
The overall trend agrees with the findings of recent work~\cite{zhang2024phase}. It is attributed to the underlying phase separation mechanism driven purely by the preferential attraction of polymer to one solvent over the other. In our work, we report this trend for three types of materials systems, including small molecule-cosolvents system (SM-S-S). 

The difference in the range of miscibility depth between material systems highlights the sensitivity of phase diagrams to the change in the interaction parameters. When the combination of interaction parameters for a given material system lands within these octants, little change is expected in the type of phase diagram and its characteristics.

\subsection{Validation of the simulated phase diagrams on typical organic thin film material systems}
In this section, we experimentally validate the existence rules reported in the previous subsections. 
The validation involves checking whether the different types of phase diagrams simulated above have been experimentally identified and whether these experimental phase diagrams can be properly simulated using available experimental data on the interaction parameters and material molar sizes. 
Phase diagrams of fully miscible ternary mixtures with all interaction parameters below the critical values (type [100]) are straightforward and will not be discussed in more detail. 
Experimental phase diagrams containing one miscibility gap are well documented, and numerous examples can be found in the literature~\cite{tan_thermodynamic_2008,liu_study_2019,saxena_studies_2002,wang_polymerization_1999,jung_polymerization_2010,tseng_interaction_1987,wolf_making_2010,gonzalez-leon_phase_2003}. 
However, material property data for more complex phase diagrams with two~\cite{lai_construction_1998} or three~\cite{imagawa_characteristics_2016,imagawa_mechanism_2016,jeong_influence_2006} miscibility gaps are rarely reported. 

The experimentally obtained phase diagrams used for the validation are sorted depending on the number of interaction parameters above the critical values, $\chi_{ij} > \chi_{ij}^c$.  This condition is used as a hard constraint for the interaction parameters used in the simulations. Still, within this constraint, some freedom is given for the choice of the exact value of the interaction parameters to compensate for the simplicity of the free energy function (Eq.~\ref{eq:FloryHuggins}) as compared to the real behavior of experimental systems. 
For instance, even without leaving the framework of the Flory-Huggins theory, the interaction parameters are often found to be significantly dependent on the volume fraction, whereas it is assumed to be constant in the theory used in the present work.
Details of the interaction parameters and the molar sizes used for the calculations can be found in the Supporting information~\textbf{SI-5}. Note that the choice of the appropriate molar sizes to be used in equation~\ref{eq:FloryHuggins}, in particular for polymers, has been a topic of discussion~\cite{van_leuken_theoretically_2024}. 

\begin{figure}[H]
    \centering
    \includegraphics[width=\textwidth,clip,trim={0cm 3.1cm 0cm 0cm}]{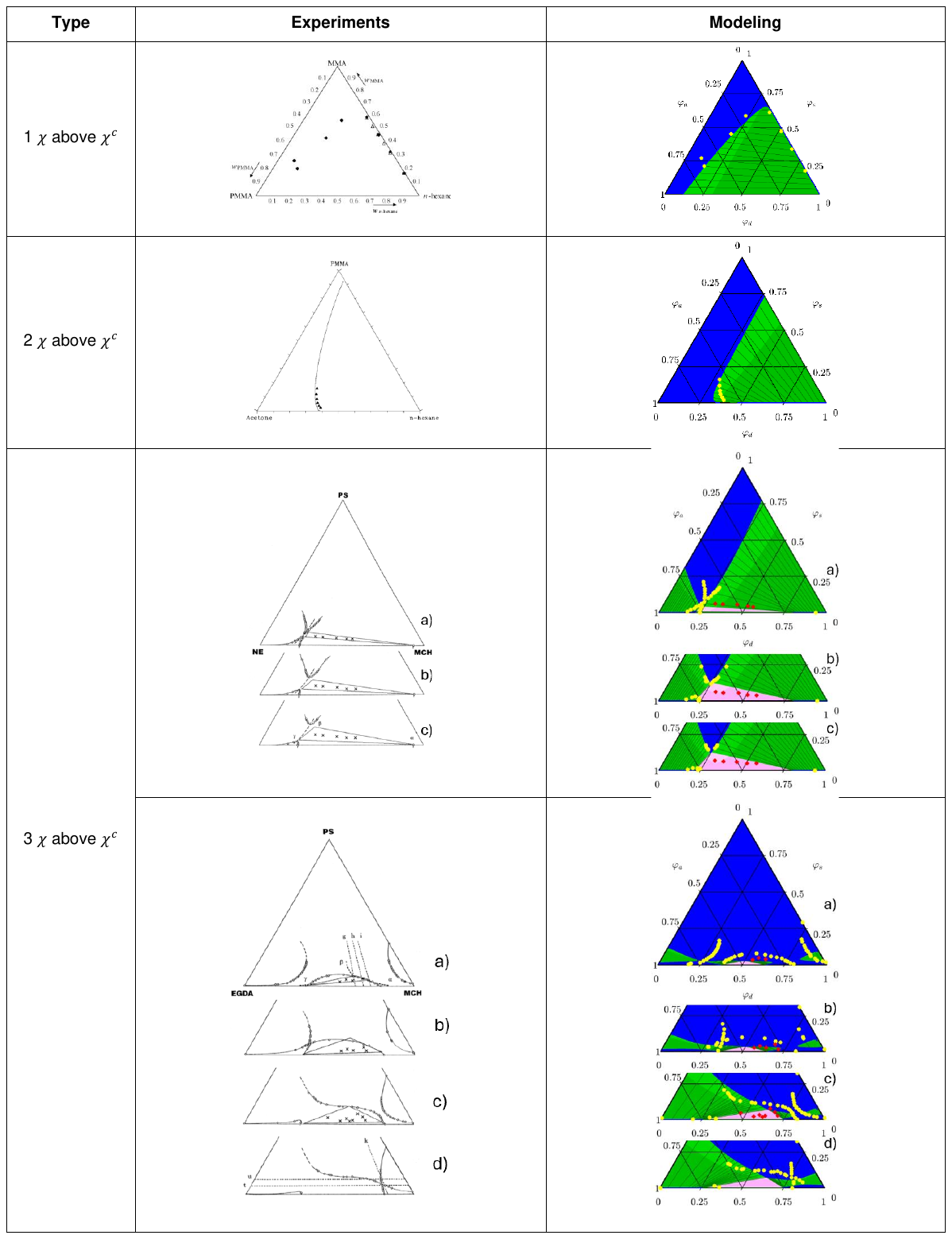}  
\caption{Comparison between experimental (left) and calculated (right) phase diagrams for three different octants of the parameter space. The experimental data are overlaid with the calculated phase diagrams in the right column. Yellow and red points correspond to the binodal curve and compositions within the three-phases equilibrium region, respectively. Reproduced with permission from Refs.~\cite{jung_polymerization_2010} ~\cite{lai_construction_1998} ~\cite{imagawa_mechanism_2016} ~\cite{imagawa_characteristics_2016}  }
    \label{fig:ModelledPhaseDiagrams}
\end{figure}

First, we use the polymer / monomer / non-solvent system PMMA (Poly(methyl 2- methylpropenoate)) /MMA (methyl 2- methylpropenoate) / n-hexane mixture studied by Jung and coworkers~\cite{jung_polymerization_2010} as a representative example of classical phase diagrams with one miscibility gap and only one interaction parameter above the critical value. The simulated phase diagrams shown in the first row of Figure~\ref{fig:ModelledPhaseDiagrams} match the experimental diagram both qualitatively and quantitatively. As expected from the design rules, the single miscibility gap in the phase diagram corresponds to a binary equilibrium region without a ternary equilibrium region (type [110]). 

Second, the polymer / solvent / non-solvent system PMMA / acetone / n-hexane mixture investigated by Lai and coworkers~\cite{lai_construction_1998} is investigated. It features two interaction parameters (PMMA/n-hexane and n-hexane/acetone) above the critical values~\textbf{(see SI-5)}. The corresponding phase diagram shown in the second row of Figure~\ref{fig:ModelledPhaseDiagrams} respects the design rules as well, with two miscibility gaps merged to form a single binary equilibrium region (type [210]). Here again, the calculated phase diagram nicely matches the experimental phase diagram. 

Third, Imagawa and coworkers measured the PS (polystyrene) / MCH (Methylcyclohexane) / NE (nitroethane) phase diagram at different temperatures~\cite{imagawa_mechanism_2016}. For this system, the three interaction parameters are above the critical values. The experimental phase diagrams (third row of Figure~\ref{fig:ModelledPhaseDiagrams}) respect the existence rules described above for this particular mixture. The three miscibility gaps merge and form a three-phase equilibrium region (type [331]) close to the bottom boundary of the phase diagram. The difference between panels a) to c) stems from the temperature change and corresponding changes in the interaction parameters. Nevertheless, the three interaction parameters remain above their critical values, the type of phase diagram remains the same, and the two- and three-phase equilibrium regions extend. This could be confirmed in the simulations, whereby a limited increase of the interaction parameters upon temperature decrease justifies the extension of the two-phase and three-phase immiscibility regions.

Fourth, Imagawa and coworkers~\cite{imagawa_characteristics_2016} investigated the temperature-dependence of the PS/ MCH / EGDA (ethylene glycol diacetate)  phase diagram (fourth row of Figure \ref{fig:ModelledPhaseDiagrams}). Here again, the three measured interaction parameters are above their critical values. All phase diagrams feature a three-phase equilibrium triangle. Remarkably, upon temperature change, Imagawa and coworkers observed a progressive transition (from (a) to (d)) from a type [151] phase diagram with 3 miscibility gaps to a type [241] phase diagram where 2 miscibility gaps merged, and finally to a type [331] phase diagram where all the miscibility gaps are merged. The calculated phase diagrams successfully mimic this transition. This nicely corresponds to the transition between phase diagrams with three interaction parameters above the critical value. As discussed above, when the interaction parameters increase, starting slightly above the critical values, the system may travel in the parameter space from the rare [151] type to the [241] type and, finally, the common [331] type.  

These results confirm the experimental relevance of the proposed existence rules, not only in terms of the type of phase diagram but also the transition between phase diagram types. The presented validation involves experimentally measured phase diagrams for several of the twenty-one different types identified in Figure~\ref{fig:zoology}, and confirms the existence rules based on the number of $\chi_{ij}$ values above the critical values $\chi_{ij}^c$. Moreover, with the material properties extracted from the literature, a quantitative and qualitative match of the phase diagram has been observed. 
For now, the validation scope is limited to only 5 types of phase diagrams, for which material properties and experimental phase diagrams can be found in the literature. However, further experimental assessment of additional phase diagrams is a topic for future prospects.

%%%%%%%%%%%%%%%%%%%%%%%%%
\section{Conclusions}
This paper introduces the taxonomy (classification) of phase diagrams in amorphous ternary systems. 
The taxonomy illustrates the diversity of phase diagrams for particular material systems using a range of interaction parameters.
Twenty-one types of phase diagrams for three amorphous ternary material systems are reported and grouped between the dominant, less common, and infrequent phase diagrams. 
The mapping of the type of phase diagram to the range of interaction parameters is provided and discussed. 
In most cases, the number of immiscible component pairs is sufficient to group the phase diagrams by type. 
However, when the binary interaction parameters are close to their critical parameters, the classification becomes very sensitive to changes in interaction parameters. 

The results reported in this paper have important implications for designing the manufacturing of flexible organic electronics - the primary science driver behind this work.
Only when one of the component pairs is immiscible, the processability and the associated miscibility depth offers a wide range of values. 
On the other hand, when two pairs of components are immiscible, changing one of the components significantly (and its interactions with two remaining components) does not change the miscibility gap (and processability), and thus will not provide significant degrees of freedom for design and tunability of device performance.

%%%%%%%%%%%%%%%%%%%%%%%%%%%%%%%%%%%%%%%%%%%%%%%%%%%%%%%%%%%%%%%%%%%%%
%% The "Acknowledgement" section can be given in all manuscript
%% classes.  This should be given within the "acknowledgement"
%% environment, which will make the correct section or running title.
%%%%%%%%%%%%%%%%%%%%%%%%%%%%%%%%%%%%%%%%%%%%%%%%%%%%%%%%%%%%%%%%%%%%%
\begin{acknowledgement}
This work was supported by the National Science Foundation (1906344) and the European Commission (H2020 Program, Project 101008701/ EMERGE) and by the German Research Foundation (DFG, Project HA 4382/14-1).
\end{acknowledgement}

%%%%%%%%%%%%%%%%%%%%%%%%%%%%%%%%%%%%%%%%%%%%%%%%%%%%%%%%%%%%%%%%%%%%%
%% The appropriate \bibliography command should be placed here.
%% Notice that the class file automatically sets \bibliographystyle
%% and also names the section correctly.
%%%%%%%%%%%%%%%%%%%%%%%%%%%%%%%%%%%%%%%%%%%%%%%%%%%%%%%%%%%%%%%%%%%%%
%\bibliographystyle{plain}
%\bibliography{achemso-demo}
\bibliography{references}

\newpage

%%%%%%%%%%%%%%%%%%%%%%%%%%%%%%%%%%%%%%%%%%%%%%%%%%%%%%%%%%%%%%%%%%%%%
%% The same is true for Supporting Information, which should use the
%% suppinfo environment.
%%%%%%%%%%%%%%%%%%%%%%%%%%%%%%%%%%%%%%%%%%%%%%%%%%%%%%%%%%%%%%%%%%%%%
\begin{suppinfo}

\section{SI-1: Additional results for the octant-based existence rules}

The reported existence rules are based on the number of immiscible pairs of components. For the design space considered in this work, three octants match this rule when one and two pairs of interaction parameters are above a critical value. In the main paper, only one octant is depicted. This part of SI provides all three octants for each material system considered. 

Figure~\ref{fig:add:oneChis} depicts the results for all octants with one interaction parameter above the critical value. Note the general trends for all three material systems, with the three-phase equilibrium region appearing when one interaction parameter is above the critical value ($\bar{\chi}>1$), while two others are slightly below the critical values ($\bar{\chi}$ slightly below 1). In some cases (Figure~\ref{fig:add:oneChis} - middle row) of the presented results, type [131] is reported for a very narrow range of interaction parameters and only for SM-S-S system. This means that type [131] is less likely when the two materials with the largest molar sizes make up the immiscible pair. Beyond this, type [131] is less common for systems with high molar size (P-SM-S) across three octants, with type [110] dominating the parameter space. 

\begin{figure}[H]
    \includegraphics[width=0.95\textwidth]{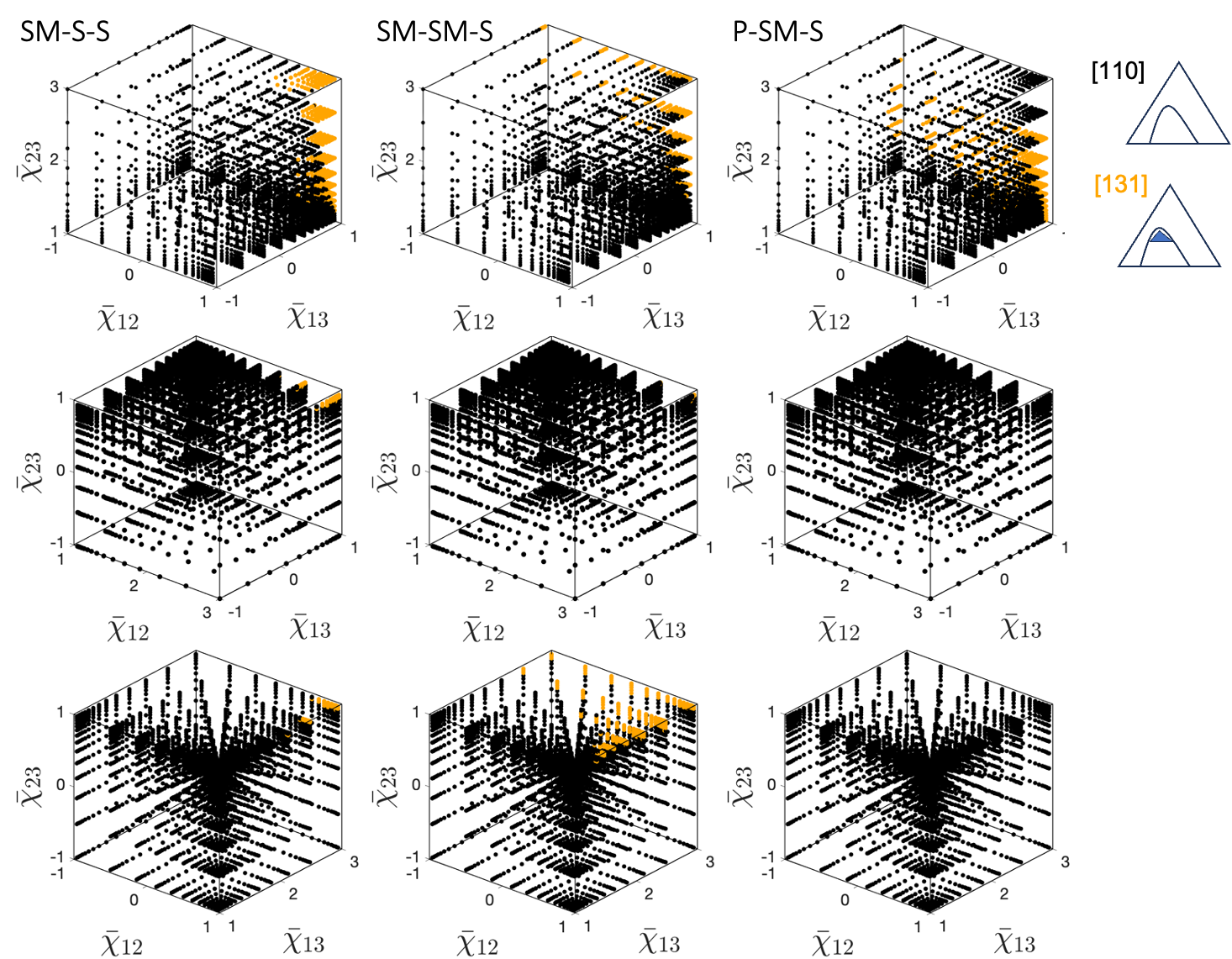}
    \caption{The distribution of the phase diagram types for all octants with one $\chi>\chi^c$: (top) $\chi_{23}$ is above the critical value, (middle) $\chi_{12}$ is above the critical value and (bottom) $\chi_{13}$ is above the critical value.}
    \label{fig:add:oneChis}
\end{figure}

Similarly, Figure~\ref{fig:add:twoChis} depicts the results for all three octants with two interaction parameters above the critical values $\chi>\chi^c$. The general trends are maintained across the three material systems studied here. Phase diagrams of type [120] with two miscibility gaps are more common when the interaction parameter of the miscible materials pair is only slightly below the critical value. Note the blue points located at the corresponding corners of the octants across all nine panels. Phase diagrams of type [231] are reported in two corners of the presented octants: the three-phase region emerges when one of the interaction parameters above the critical value becomes very large. The central panel of Figure~\ref{fig:add:twoChis} showcases this situation. The deep blue points are located in two corners of the octant, depending on the origin of the three-phase regions. When the value of $\bar{\chi_{23}}$ is high, the three-phase region links this interaction parameter, while when $\bar{\chi_{13}}$ is high, the three-phase region links to this interaction parameter.

\begin{figure}[H]
    \includegraphics[width=0.85\textwidth]{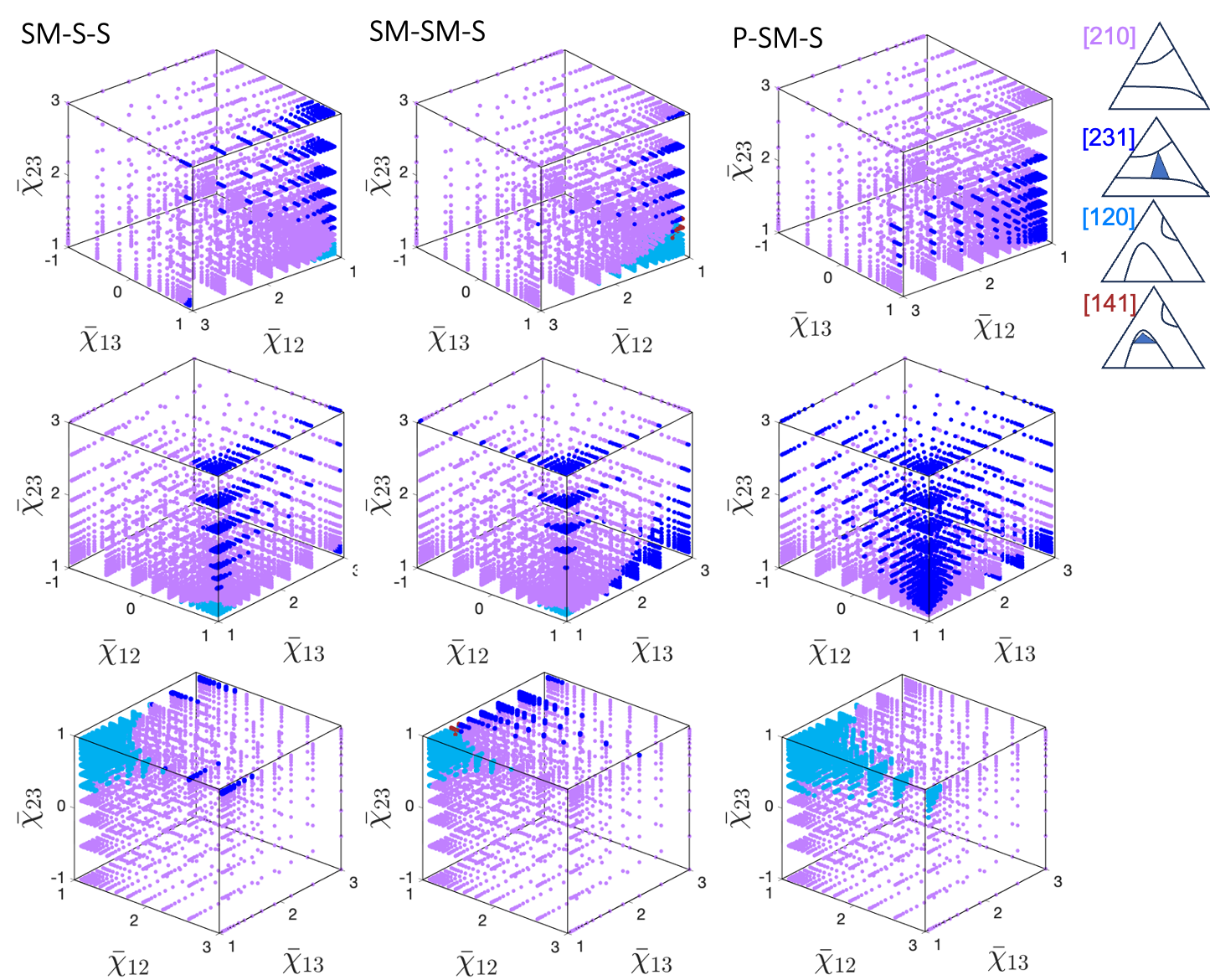}
    \caption{The distribution of the phase diagram types for all octants with two $\chi>\chi^c$.}
    \label{fig:add:twoChis}
\end{figure}

%%%%%%%%%%%%%%%%%%%%%%%%%%%%%%%%%%%
\section{SI-2: Distribution of the phase diagram types at the selected planes for the design space}
Figure~\ref{fig:add:oneChisXsec} shows phase diagrams on a plane of the parameter space, corresponding to a fixed value of $\chi_{12}$ slightly below the critical value ($\chi_{12}=0.12408<\chi_{12}^c$) and a selected range for the two remaining interaction parameters. One of the interaction parameters is below its critical value ($\chi_{13}<\chi_{13}^c$) while the other is above the critical value ($\chi_{23}>\chi_{23}^c$). 
As $\chi_{13}$ approaches its critical value and $\chi_{23}$ is slightly above the critical value (right bottom corner of the figure), the two-phase region is significantly smaller than when $\chi_{13}$ is significantly lower (left bottom corner of the figure). This is a typical counter-intuitive situation where increasing the miscibility of one material pair decreases the miscibility of the ternary blend. We attribute this behavior to the underlying mechanism of phase separation, similar to one reported by~\cite{zhang2024phase} for the closed-loop isolate miscibility gap. When $\chi_{13}$ is significantly lower than the critical value, the phase separation is dominated by the preferential attraction of components 1 and 3 over the other (2 and 3). As $\chi_{13}$ increases and approaches the critical value, the three interactions become stronger, resulting in small miscibility gaps. However, this requires additional analytical analysis that we defer to a separate publication.

\begin{figure}[H]
    \includegraphics[width=0.65\textwidth]{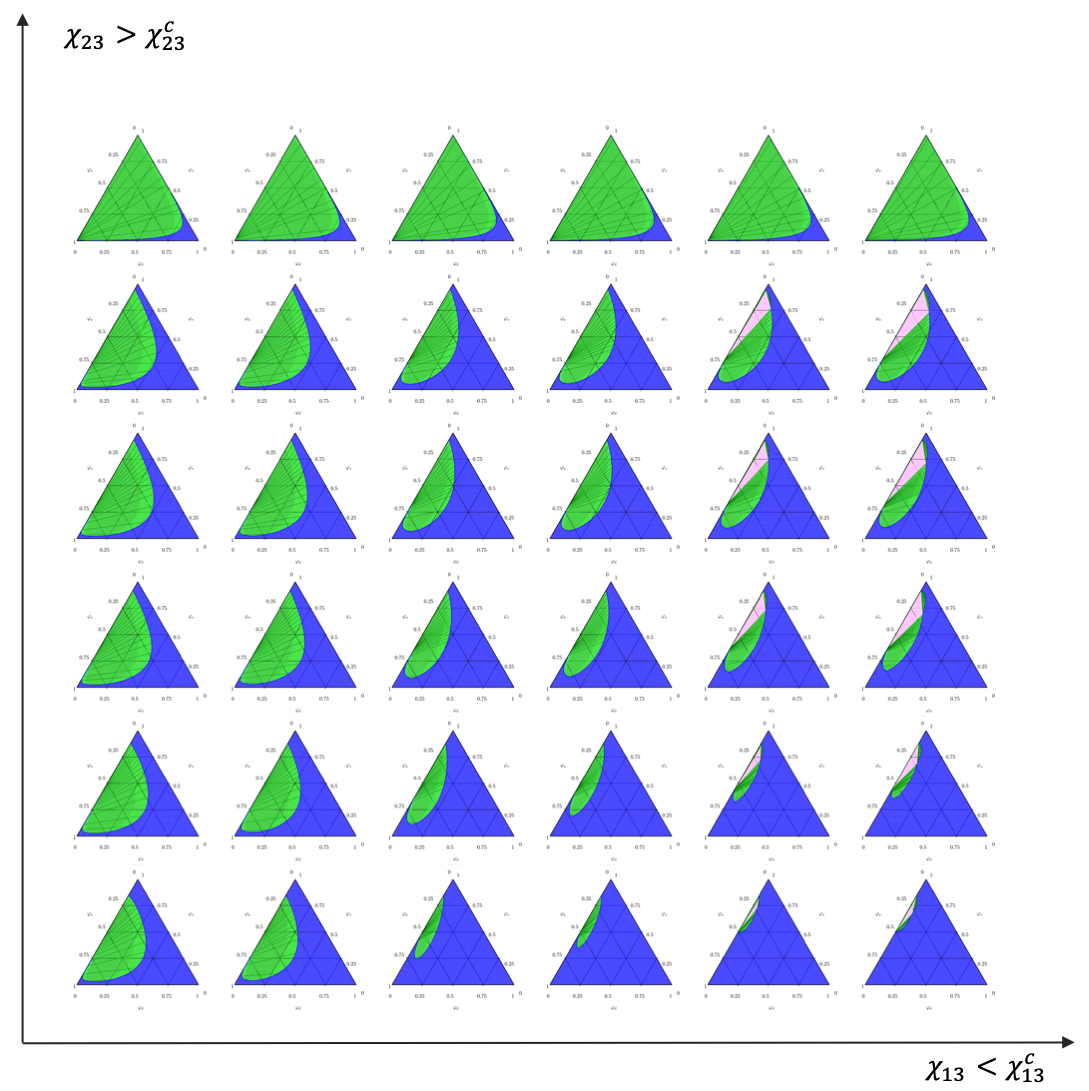}
    \caption{The distribution of the phase diagram types at the selected plane $\chi_{12}=0.12408$ ($<\chi_{12}^c$), when one $\chi>\chi^c$. The phase diagrams included in the figure correspond to the following:     
    $\chi_{13}= \{0.052495, 0.17143, 0.46028, 0.48475, 0.5291, 0.53763 \}$
    $\chi_{23}= \{1.0996, 1.1359, 1.1974 ,1.2427, 1.3017, 3.1416 \}$.
    }
    \label{fig:add:oneChisXsec}
\end{figure}

Similarly, Figure~\ref{fig:add:twoChisXsec} showcases the distribution of phase diagrams when two interaction parameters are above the critical values. The phase diagrams are selected for a fixed value of $\chi_{13}$ slightly below the critical value ($\chi_{13}=0.53763<\chi_{13}^c$).The two other interaction parameters vary above the critical value (as noted in the figure). Note the non-trivial distribution of phase diagram types. The type [141] can be found only when three interaction parameters are close to their critical values (left bottom corner). As $\chi_{23}$ increases, the two miscibility gaps merge (type [231]) and become larger as expected. Still, then the three-phase regions disappear, as the tie-lines for two two-phase regions align with each other and merge into the two-phase region (see the first column of the figure). This trend is consistent across all the columns in this figure. However, the dominance of the three-phase region depends non-linearly on the relative values of the interaction parameters - as depicted in Figure~\ref{fig:add:oneChis}.

\begin{figure}[H]
    \includegraphics[width=0.65\textwidth]{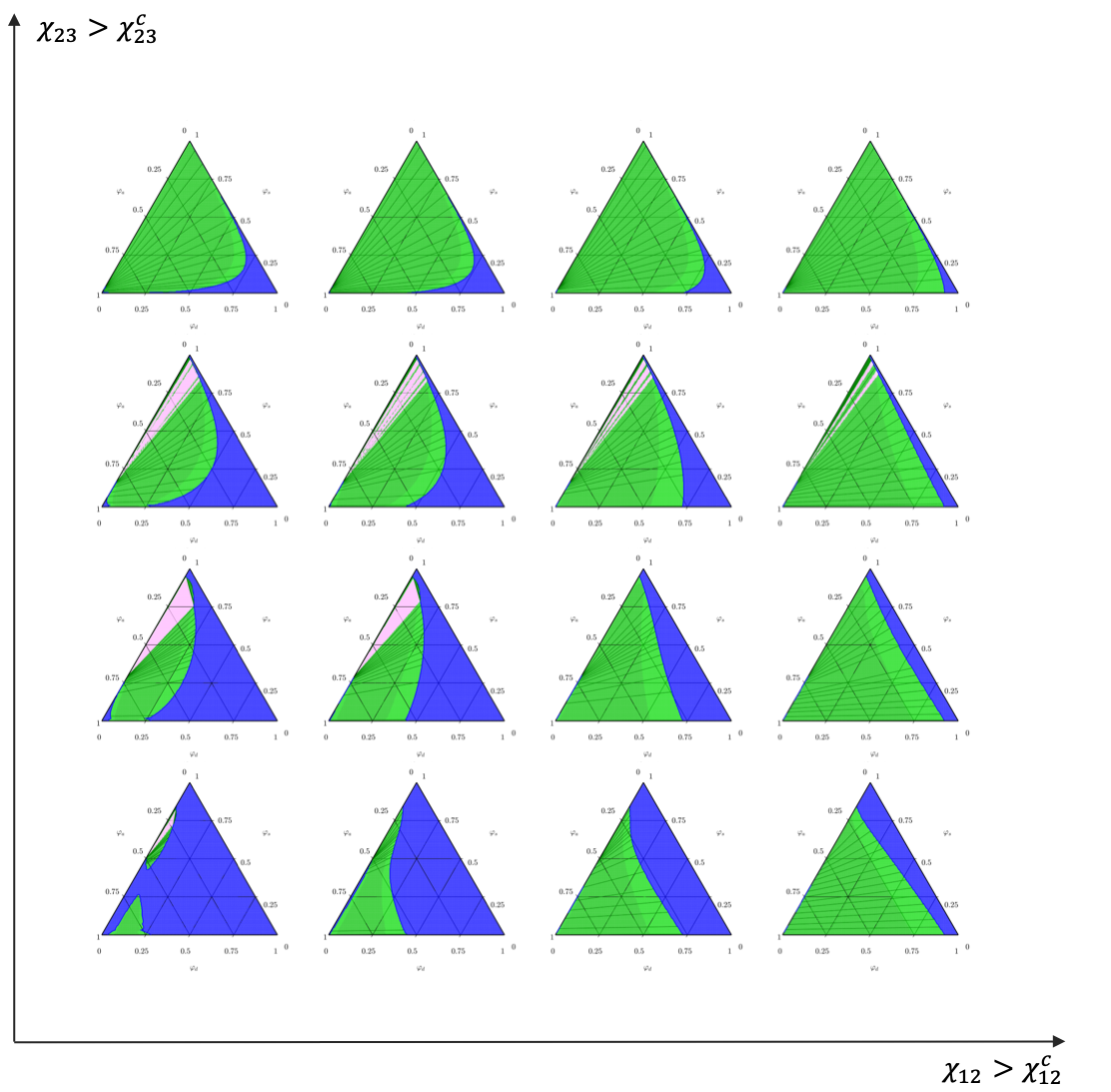}
    \caption{The distribution of the phase diagram types at the selected plane $\chi_{13}=0.53763$, when two $\chi>\chi^c$. The phase diagrams included in the figure correspond to the following:  $\chi_{12}= \{0.13714,0.155,0.22166,0.39184\}$, 
    $\chi_{23}=\{1.0996, 1.3017,1.6081,2.6565\}$. 
    }
    \label{fig:add:twoChisXsec}
\end{figure}

\section{SI-3: Features of phase diagrams}
Figure~\ref{fig:sensitivity} depicts the definition of the miscibility depth along three composition directions for types [110] and [210]. There are three panels for each type of phase diagram, each panel corresponding to the permutations of the interaction parameters.

\begin{figure}[H]
    \includegraphics[width=0.95\textwidth]{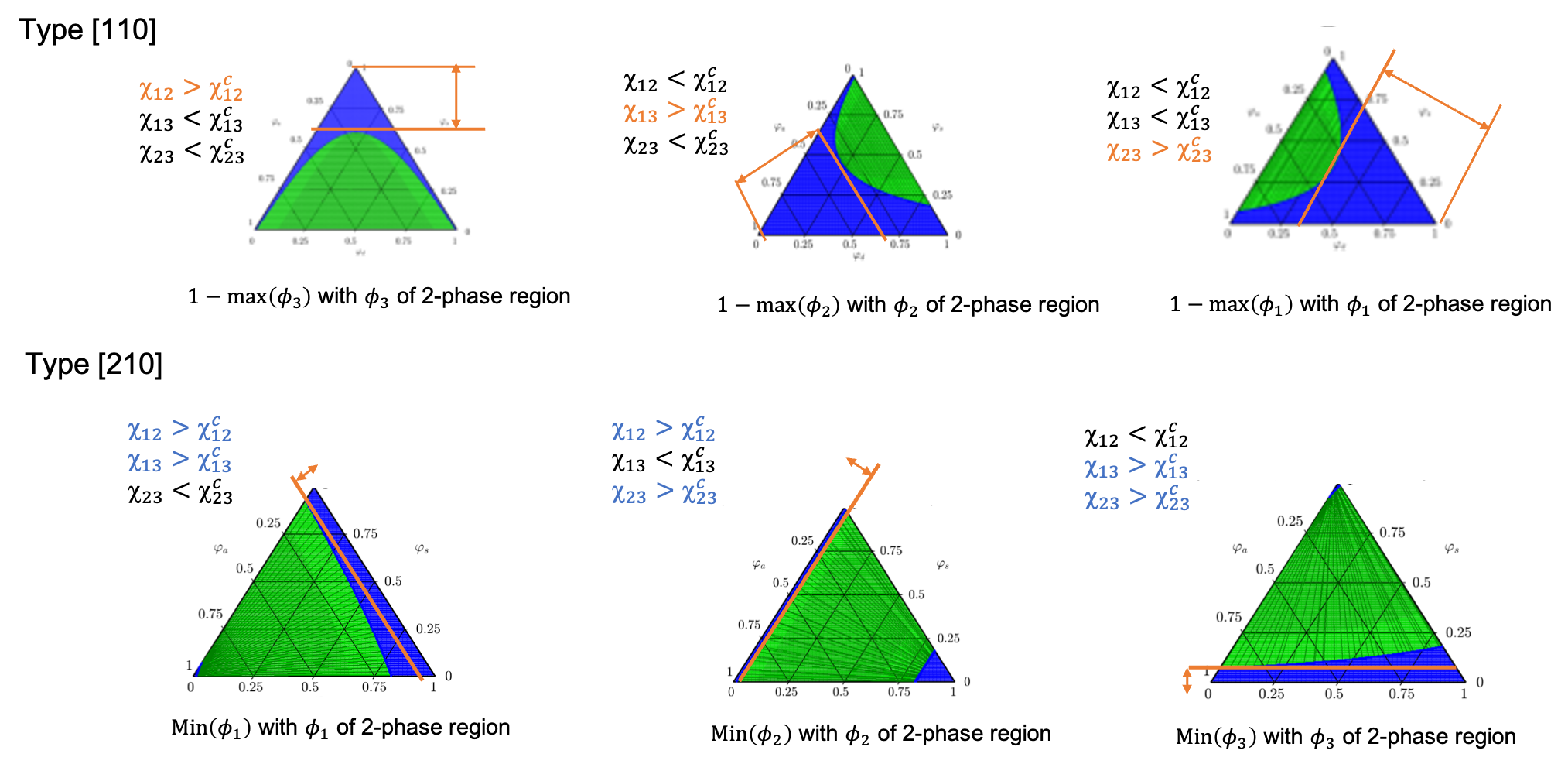}
    \caption{Characteristic quantities extracted from phase diagrams for type [110] (top) and [210] (bottom).}
    \label{fig:cut}
\end{figure}

 \section{SI-4: Parameters used for each type of phase diagram identified}
 The following table gives the material parameters used to reproduce the different types of phase diagrams identified in the main text. The density is fixed to $1000 \ kg \ m^{-3}$ for all the materials.
 \begin{table}[H]

\begin{center}
\begin{tabular}{
|p{0.1\textwidth}
|p{0.3\textwidth}
|p{0.5\textwidth}|}
\hline 
\textbf{Phase diagram type} & \textbf{Molar size}& \textbf{Interaction parameters}  \\
\hline 
$[100]$ & $N_{1}=245, N_{2}=5, N_{3}=1$ & $\chi_{12}=0.0121, \chi_{13}=0.0525, \chi_{23}=0.0971$ \\
\hline
$[110-o]$ & $N_{1}=5, N_{2}=1, N_{3}=1$ & $\chi_{12}=1.0472, \chi_{13}=0.9949, \chi_{23}=1.8699$ \\
\hline
$[110]$ & $N_{1}=5, N_{2}=1, N_{3}=1$ & $\chi_{12}=2.2837, \chi_{13}=1.0472, \chi_{23}=1.8699$ \\
\hline
$[131]$ & $N_{1}=5, N_{2}=1, N_{3}=1$ & $\chi_{12}=2.6565, \chi_{13}=0.897, \chi_{23}=1.8699$ \\
\hline
$[120]$ & $N_{1}=5, N_{2}=1, N_{3}=1$ & $\chi_{12}=1.1974, \chi_{13}=0.8517, \chi_{23}=2.1694$ \\
\hline
$[210]$ & $N_{1}=5, N_{2}=5, N_{3}=1$ & $\chi_{12}=0.42, \chi_{13}=0.3172, \chi_{23}=1.3784$ \\
\hline
$[141]$ & $N_{1}=5, N_{2}=5, N_{3}=1$ & $\chi_{12}=0.4339, \chi_{13}=0.9949, \chi_{23}=1.4782$ \\
\hline
$[231]$ & $N_{1}=5, N_{2}=1, N_{3}=1$ & $\chi_{12}=1.1974, \chi_{13}=0.9791, \chi_{23}=3.8145$ \\
\hline
$[162]$ & $N_{1}=1, N_{2}=1, N_{3}=245$ & $\chi_{12}=1.999, \chi_{13}=1.0649, \chi_{23}=1.0649$ \\
\hline
$[130]$ & $N_{1}=5, N_{2}=1, N_{3}=1$ & $\chi_{12}=1.3017, \chi_{13}=1.2427, \chi_{23}=2.2204$ \\
\hline
$[151]$ & $N_{1}=5, N_{2}=1, N_{3}=1$ & $\chi_{12}=1.3017, \chi_{13}=1.3017, \chi_{23}=2.2869$ \\
\hline
$[172]$ & $N_{1}=1, N_{2}=1, N_{3}=1$ & $\chi_{12}=2.6, \chi_{13}=2.6, \chi_{23}=2.6$ \\
\hline
$[193]$ & $N_{1}=1, N_{2}=1, N_{3}=1$ & $\chi_{12}=2.65, \chi_{13}=2.65, \chi_{23}=2.65$ \\
\hline
$[220]$ & $N_{1}=5, N_{2}=1, N_{3}=1$ & $\chi_{12}=1.1626, \chi_{13}=2.2837, \chi_{23}=2.6324$ \\
\hline
$[241]$ & $N_{1}=5, N_{2}=5, N_{3}=1$ & $\chi_{12}=0.4441, \chi_{13}=1.0996, \chi_{23}=1.6091$ \\
\hline
$[262]$ & $N_{1}=1, N_{2}=1, N_{3}=1$ & $\chi_{12}=2.7, \chi_{13}=2.72, \chi_{23}=2.63$ \\
\hline
$[283]$ & $N_{1}=1, N_{2}=1, N_{3}=1$ & $\chi_{12}=2.7, \chi_{13}=2.7, \chi_{23}=2.65$ \\
\hline
$[331]$ & $N_{1}=1, N_{2}=1, N_{3}=1$ & $\chi_{12}=2.1694, \chi_{13}=2.3734, \chi_{23}=3.8145$ \\
\hline
$[352]$ & $N_{1}=1, N_{2}=1, N_{3}=1$ & $\chi_{12}=2.63, \chi_{13}=2.75, \chi_{23}=2.8$ \\
\hline
$[373]$ & $N_{1}=1, N_{2}=1, N_{3}=1$ & $\chi_{12}=2.67, \chi_{13}=2.7, \chi_{23}=2.65$ \\
\hline
$[463]$ & $N_{1}=5, N_{2}=1, N_{3}=1$ & $\chi_{12}=1.6081, \chi_{13}=1.6081, \chi_{23}=2.2204$ \\
\hline

\end{tabular}
\caption{Material parameters to illustrate the different types of phase diagrams identified}
\label{tab:ParamsperType}
\end{center}
\end{table}

\section{SI-5: Parameters used for the modeling of experimental phase diagrams}
For the different systems studied, the molar sizes $N_{i}=\frac{v_{i}}{v_{0}}$ are calculated using the species molar volumes $v_{i}=\frac{M_{i}}{\rho_{i}}$, where $M_{i}$ and $\rho_{i}$ are the species molar mass and density available from literature data, respectively. The values used for all studied material systems are summarized in Table~\ref{tab:MaterialParams}. Note that the temperature dependence of the density is taken into account for the PS/NE/MCH and PS/EGDA/MCH mixtures.
\begin{table}[H]

\begin{center}
\begin{tabular}{
|p{0.14\textwidth}
|p{0.31\textwidth}
|p{0.28\textwidth}
|p{0.18\textwidth}|}
\hline 
\textbf{Mixture} &\textbf{Density ($kg ~ m^{-3}$)} & \textbf{Molar mass ($kg ~ mol^{-1}$)} & \textbf{Molar size} \\
\hline 

\text{n-hexane} (1)&
$\rho_{1}=613.4 \ [\cite{jung_polymerization_2010}]$ & $M_{1}=86.18*10^{-3}$ & $N_{1}=1.24$\\
\cline{2-4}
\text{PMMA(2)} &$\rho_{2}=1153$ \ [\cite{jung_polymerization_2010}]& $M_{2}=50 \ [\cite{jung_polymerization_2010}]$ & $N_{2}=384$\\
\cline{2-4}
\text{MMA (3)}&$\rho_{3}=886.8 \ [\cite{jung_polymerization_2010}]$ & $M_{3}=100.121*10^{-3}$ & $N_{3}=1$\\
\hline
\text{n-hexane(1)} &$\rho_{1}=613.4 \ [\cite{lai_construction_1998}]$ & $M_{1}=86.18*10^{-3}$ & $N_{1}=1.90$\\
\cline{2-4}
\text{acetone (2)} &$\rho_{2}=784$ \ [\cite{lai_construction_1998}]& $M_{2}=58.08*10^{-3}$ & $N_{2}=1$\\
\cline{2-4}
\text{PMMA (3)} &$\rho_{3}=1153 \ [\cite{lai_construction_1998}]$ & $M_{3}=159.97 \ [\cite{lai_construction_1998}]$ & $N_{3}=1873$\\
\hline
\text{MCH} (1)&$\rho_{1}=786.36-0.8479*10^{-1}*T-3.75*10^{-4}*T^{2}$ \ [\cite{imagawa_mechanism_2016}] & $M_{1}=98.186*10^{-3}$ & $N_{1}=1$\\
\cline{2-4}
\text{NE} (2)&$\rho_{2}=1073.06-1.1922*10^{-1}*T-3.65*10^{-4}*T^{2} $ \ [\cite{imagawa_mechanism_2016}]& $M_{2}=75.07*10^{-3}$ & $N_{2}=1.09$\\
\cline{2-4}
\text{PS} (3)&$\rho_{3}=1086.5-6.19*10^{-1}*T+1.36*10^{-4}*T^{2}$ \ [\cite{imagawa_mechanism_2016}] & $M_{3}=96.4 \ [\cite{imagawa_mechanism_2016}]$ & $N_{3}=\frac{981.81*\rho_{1}(T)}{\rho_{3}(T)}$\\
\hline
\text{MCH} (1)&$\rho_{1}=769.37 \ [\cite{imagawa_characteristics_2016}]$ & $M_{1}=98.186*10^{-3}$ & $N_{1}=1$\\
\cline{2-4}
\text{EGDA} (2)&$\rho_{2}=1106.3$ \ [\cite{imagawa_characteristics_2016}]& $M_{2}=146.142*10^{-3}$ & $N_{2}=1.04$\\
\cline{2-4}
\text{PS} (3)&$\rho_{3}=1086.5-6.19*10^{-1}*T+1.36*10^{-4}*T^{2} \ [\cite{imagawa_characteristics_2016}]$ & $M_{3}=48 \ [\cite{imagawa_characteristics_2016} $] & $N_{3}=\frac{3.76*10^{5}}{\rho_{3}(T)}$\\
\hline
\end{tabular}
\caption{Material systems properties}
\label{tab:MaterialParams}
\end{center}
\end{table}
\noindent
\subsection{PMMA/MMA/n-hexane phase diagram}

\noindent
The interaction parameter values $\chi_{ij}$ measured by Jung and coworkers~\cite{jung_polymerization_2010} are used for the phase diagram calculation (first row and last column of Figure~\ref{fig:ModelledPhaseDiagrams}). The critical interaction parameters $\chi_{ij}^c$ are calculated using equation \ref{eq:CriticalInteractionParameter}. For this studied system, one interaction parameter is above the critical values (See Table~\ref{tab:Type1InteractionParam}).
\end{suppinfo}
\begin{table}[H]

\begin{center}
\begin{tabular}{
|p{0.3\textwidth}
|p{0.4\textwidth}|}
\hline 
\textbf{Interaction parameter} & \textbf{Critical interaction parameter}  \\
\hline
$\chi_{12}=1.31 \ [\cite{jung_polymerization_2010}]$ & $\chi_{12}^c=0.45$ \\
\hline
$\chi_{13}=1.49 \ [\cite{jung_polymerization_2010}]$ & $\chi_{13}^c=1.8$\\
\hline
$\chi_{23}=0.412 \ [\cite{jung_polymerization_2010}]$ & $\chi_{23}^c=0.55$ \\
\hline

\end{tabular}
\caption{n-hexane (1), PMMA (2),  MMA (3) interaction parameters}
\label{tab:Type1InteractionParam}
\end{center}
\end{table}

\subsection{PMMA/acetone/n-hexane phase diagram}

\noindent
The interaction parameters are determined experimentally with volume fraction dependency by Lai and coworkers~\cite{lai_construction_1998} and are shown in the first column of Table~\ref{tab:Type2InteractionParam}. For the whole composition space, two interaction parameters are above the critical values (compare columns 1 and 3 of Table~\ref{tab:Type2InteractionParam}). Using this constraint, the PMMA/acetone/n-hexane phase diagram is calculated (second row and last column of Figure~\ref{fig:ModelledPhaseDiagrams}) using the interaction parameters values shown in the second column of Table~\ref{tab:Type2InteractionParam}.
\begin{table}[H]

\begin{center}
\begin{tabular}{
|p{0.3\textwidth}
|p{0.3\textwidth}
|p{0.3\textwidth}|}
\hline 
\textbf{Interaction parameter from experiments} & \textbf{Interaction parameter for modeling}& \textbf{Critical interaction parameter}  \\
\hline 
$\chi_{12}^{exp}=1.574+\frac{2.277}{1+3.026\varphi_{2}} \ [\cite{lai_construction_1998}] $ & $\chi_{12}^{mod}=1.495$ & $\chi_{12}^c=1.49$ \\
\hline
$\chi_{13}^{exp}=2.07 \ [\cite{lai_construction_1998}]$ & $\chi_{13}^{mod}=0.58$ & $\chi_{13}^c=0.28$\\
\hline
$\chi_{23}^{exp}=0.25+0.12\varphi_{3} \ [\cite{lai_construction_1998}]$ & $\chi_{23}^{mod}=0.37$ & $\chi_{23}^c=0.52$ \\
\hline

\end{tabular}
\caption{n-hexane (1), PMMA (2), MMA (3) interaction parameters}
\label{tab:Type2InteractionParam}
\end{center}
\end{table}
\subsection{PS/MCH/EGDA phase diagram}

\noindent
At the measured phase diagrams temperatures (32.9°C, 33°C, 34°C, and 34.8 °C), the interaction parameters evaluated using Hildebrand solubilities are given in Table~\ref{tab:Type3HildebrandInteraction}. The critical interaction parameters measured experimentally using Hildebrand solubilities and the binodal critical temperatures~\cite{imagawa_characteristics_2016} are $\chi_{12}^{c,exp}=1.66$, $\chi_{13}^{c,exp}=0.54$, and $\chi_{23}^{c,exp}=0.27$.  Three interaction parameters are above the critical values for the four considered temperatures. The PS/MCH/EGDA phase diagrams are calculated (fourth row and last column of Figure~\ref{fig:ModelledPhaseDiagrams}) at different temperatures respecting this constraint but for the critical interaction parameters obtained from the molar sizes given in Table~\ref{tab:MaterialParams} ($\chi_{12}^{c}=1.97$, $\chi_{13}^{c}=0.55$, and $\chi_{23}^{c}=0.54$). The interaction parameter values used for the calculation are given in Table~\ref{tab:Type3Interaction}.
\begin{table}[H]

\begin{center}
\begin{tabular}{
|p{0.5\textwidth}
|p{0.1\textwidth}
|p{0.1\textwidth}
|p{0.1\textwidth}
|p{0.1\textwidth}|}
\hline 
\diagbox{Interaction parameter}{Temperature} & \textbf{32.9°C} & \textbf{33°C} & \textbf{34°C}& \textbf{34.8°C}\\
\hline 
$\chi_{12}^{exp}$ \ [\cite{imagawa_characteristics_2016}] & $1.6867$ & $1.6861$ & $1.6806$ & $1.6763$ \\
\hline
$\chi_{13}^{exp}$ \ [\cite{imagawa_characteristics_2016}]& $0.5460$ & $0.5458$ & $0.5441$ & $0.5426$\\
\hline
$\chi_{23}^{exp}$ \ [\cite{imagawa_characteristics_2016}] & $0.3134$ & $0.3133$ & $0.3122$ & $0.3114$ \\
\hline

\end{tabular}
\caption{Measured interaction parameters from experiments for the MCH (1), EGDA (2), PS (3) mixture}
\label{tab:Type3HildebrandInteraction}
\end{center}
\end{table}
\begin{table}[H]

\begin{center}
\begin{tabular}{
|p{0.5\textwidth}
|p{0.1\textwidth}
|p{0.1\textwidth}
|p{0.1\textwidth}
|p{0.1\textwidth}|}
\hline 
\diagbox{Interaction parameter}{Temperature} & \textbf{32.9°C} & \textbf{33°C} & \textbf{34°C}& \textbf{34.8°C}\\
\hline 
$\chi_{12}^{mod}$  & $2.15$ & $2.15$ & $2.0243$ & $2.0129$ \\
\hline
$\chi_{13}^{mod}$ & $0.567$ & $0.562$ & $0.563$ & $0.56471$\\
\hline
$\chi_{23}^{mod}$  & $0.7$ & $0.7$ & $0.549$ & $0.548$ \\
\hline

\end{tabular}
\caption{Interaction parameters used for modeling the MCH (1), EGDA (2), PS (3) phase diagrams }
\label{tab:Type3Interaction}
\end{center}
\end{table}

\subsection{PS/MCH/NE phase diagram}

\noindent
Similar to PS/MCH/EGDA, the critical interaction parameters measured experimentally using Hildebrand solubilities and the binodal critical temperatures~\cite{imagawa_mechanism_2016} are $\chi_{12}^{c,exp}=3.08$, $\chi_{13}^{c,exp}=0.51$, and $\chi_{23}^{c,exp}=0.96$. At the measured phase diagrams temperatures (2.5°C, 3°C and 3.3°C), the interaction parameters using Hildebrand solubilities are available in Table~\ref{tab:Type4HildebrandInteraction}. Three interaction parameters are above the critical values for the three considered temperatures. Here again, the phase diagrams are calculated (third row and last column of Figure~\ref{fig:ModelledPhaseDiagrams}) at different temperatures respecting this constraint, but for the critical interaction parameters obtained from the molar sizes given in Table~\ref{tab:MaterialParams} ($\chi_{12}^{c}=1.92$, $\chi_{13}^{c}=0.54$, and $\chi_{23}^{c}=0.50$). The interaction parameter values used for the calculation are given in Table~\ref{tab:Type4Interaction}.
\begin{table}[H]

\begin{center}
\begin{tabular}{
|p{0.5\textwidth}
|p{0.1\textwidth}
|p{0.1\textwidth}
|p{0.1\textwidth}|}
\hline 
\diagbox{Interaction parameter}{Temperature} & \textbf{2.5°C} & \textbf{3°C} & \textbf{3.3°C}\\
\hline 
$\chi_{12}^{exp}$ \ [\cite{imagawa_mechanism_2016}] &$3.3144$ & $3.3086$ & $3.3051$  \\
\hline
$\chi_{13}^{exp}$ \ [\cite{imagawa_mechanism_2016}] &$0.5923$  & $0.5922$ & $0.5916$  \\
\hline
$\chi_{23}^{exp}$ \ [\cite{imagawa_mechanism_2016}] & $1.1032$ & $1.1012$ & $1.1001$  \\
\hline

\end{tabular}
\caption{Interaction parameters from experiments for the MCH (1), NE (2), PS (3) mixture}
\label{tab:Type4HildebrandInteraction}
\end{center}
\end{table}
\begin{table}[H]

\begin{center}
\begin{tabular}{
|p{0.5\textwidth}
|p{0.1\textwidth}
|p{0.1\textwidth}
|p{0.1\textwidth}|}
\hline 
\diagbox{Interaction parameter}{Temperature} & \textbf{2.5°C} & \textbf{3°C} & \textbf{3.3°C}\\
\hline 
$\chi_{12}^{mod}$  & $2.2$ & $2.2$ & $2.2$  \\
\hline
$\chi_{13}^{mod}$  & $1.2$ & $1.2$ & $1.2$ \\
\hline
$\chi_{23}^{mod}$  & $0.65$ & $0.65$ & $0.59$  \\
\hline

\end{tabular}
\caption{Interaction parameters used for modeling the MCH (1), NE (2), PS (3) phase diagrams}
\label{tab:Type4Interaction}
\end{center}
\end{table}

\end{document}